\PassOptionsToPackage{dvipsnames}{xcolor}
\documentclass[11pt]{article}
\usepackage{jheppub}
\usepackage{bm,bbm}
\usepackage{booktabs,amsmath}
\usepackage{mathtools}
\usepackage{graphics}
\usepackage{braket}
\usepackage{tikz}
\usepackage{ascmac}
\usepackage{enumitem}
\usepackage{comment}
\usepackage{stackrel}
\usepackage{accents}
\usepackage{framed}

\usetikzlibrary{patterns}
\usetikzlibrary{decorations.markings}
\usetikzlibrary{decorations.pathmorphing}
\tikzset{snake it/.style={decorate, decoration=snake}}

\tikzset{test/.style n args={3}{
    postaction={
    decorate,
    decoration={
    markings,
    mark=between positions 0 and \pgfdecoratedpathlength step 0.5pt with {
    \pgfmathsetmacro\myval{multiply(
        divide(
        \pgfkeysvalueof{/pgf/decoration/mark info/distance from start}, \pgfdecoratedpathlength
        ),
        100
    )};
    \pgfsetfillcolor{#3!\myval!#2};
    \pgfpathcircle{\pgfpointorigin}{#1};
    \pgfusepath{fill};}
}}}}

\definecolor{c1}{rgb}{0.0,0.2,0.9}
\definecolor{c2}{rgb}{0.45,0.0,0.45}
\definecolor{c3}{rgb}{0.9,0.2,0.0}

\usetikzlibrary{arrows}

\makeatletter
\newcommand{\mylabel}[2]{#2\def\@currentlabel{#2}\label{#1}}
\makeatother

\newcommand{\RomanNumeralCaps}[1]
    {\MakeUppercase{\romannumeral #1}}

\usetikzlibrary{quotes,angles}

\usetikzlibrary{shadings}

\tikzset{middlearrow/.style={
        decoration={markings,
            mark= at position 0.57 with {\arrow{#1}} ,
        },
        postaction={decorate}
    }
}

\usepackage{blkarray}

\usepackage{nicematrix}
\NiceMatrixOptions%
{code-for-first-row = \scriptstyle,
code-for-first-col = \scriptstyle }

\usetikzlibrary{decorations.markings}
\def\MarkLt{4pt}
\def\MarkSep{2pt}

\tikzset{
  TwoMarks/.style={
    postaction={decorate,
      decoration={
        markings,
        mark=at position #1 with
          {
              \begin{scope}[xslant=0.2]
              \draw[line width=\MarkSep,white,-] (0pt,-\MarkLt) -- (0pt,\MarkLt) ;
              \draw[-] (-0.5*\MarkSep,-\MarkLt) -- (-0.5*\MarkSep,\MarkLt) ;
              \draw[-] (0.5*\MarkSep,-\MarkLt) -- (0.5*\MarkSep,\MarkLt) ;
              \end{scope}
          }
       }
    }
  },
  TwoMarks/.default={0.5},
  OneMark/.style={
    postaction={decorate,
      decoration={
        markings,
        mark=at position #1 with
          {
              \draw[-] (0,-\MarkLt) -- (0,\MarkLt) ;
          }
       }
    }
  },
  OneMark/.default={0.5}
}

\edef\restoreparindent{\parindent=\the\parindent\relax}
\usepackage{parskip}
\restoreparindent
\usepackage{ascmac}
\usepackage{mathrsfs}
\usepackage{enumitem}
\newlist{steps}{enumerate}{1}
\setlist[steps, 1]{label = Step \arabic*:}

\renewcommand*\arraystretch{1.2}

\def\d{{\rm d}}

\def\i{{\rm i}}

\def\CD{{\cal D}}

\def\CJ{{\cal J}}

\def\CR{{\cal R}}

\def\CO{{\cal O}}

\def\b0{\bm{0}_\perp}

\usepackage{centernot}
\usepackage{mathtools}
\usepackage{stmaryrd}

\makeatletter
\newcommand{\xMapsto}[2][]{\ext@arrow 0599{\Mapstofill@}{#1}{#2}}
\def\Mapstofill@{\arrowfill@{\Mapstochar\Relbar}\Relbar\Rightarrow}
\makeatother

\usepackage{colortbl}
\definecolor{shadecolor}{rgb}{0.90,0.90,0.90}

\newtheorem{axiom_}{Axiom}
{\begin{axiom_}\begin{shaded}}%
{\end{shaded}\end{axiom_}}

\DeclareFontFamily{U}{mathx}{\hyphenchar\font45}
\DeclareFontShape{U}{mathx}{m}{n}{
      <5> <6> <7> <8> <9> <10>
      <10.95> <12> <14.4> <17.28> <20.74> <24.88>
      mathx10
      }{}
\DeclareSymbolFont{mathx}{U}{mathx}{m}{n}
\DeclareFontSubstitution{U}{mathx}{m}{n}
\DeclareMathAccent{\widecheck}{0}{mathx}{"71}
\title{The epsilon expansion of the O$(N)$ model with line defect from conformal field theory}

\author[a]{Tatsuma Nishioka,}
\author[a,b]{Yoshitaka Okuyama}
\author[a]{and Soichiro Shimamori}

\affiliation[a]{
Department of Physics, Osaka University,\\
Machikaneyama-Cho 1-1, Toyonaka 560-0043, Japan
}

\affiliation[b]{Department of Physics, Faculty of Science,
The University of Tokyo,\\
Bunkyo-Ku, Tokyo 113-0033, Japan}

\preprint{OU-HET-1161}

\abstract{We employ the axiomatic framework
of Rychkov and Tan to investigate the critical O$(N)$ vector model with a line defect in $(4-\epsilon)$ dimensions.
We assume the fixed point is described by defect conformal field theory and show that the critical value of the defect coupling to the bulk field is uniquely fixed without resorting to diagrammatic calculations.
We also study various defect localized operators by the axiomatic method, where the analyticity of correlation functions plays a crucial role in determining the conformal dimensions of defect composite operators.
In all cases, including operators with operator mixing, we reproduce the leading anomalous dimensions obtained by perturbative calculations.
}

\begin{document}
\maketitle

\section{Introduction}
Defects such as impurities and domain walls are ubiquitous in the real world.
In particle physics, the Wilson and 't Hooft loop operators, prominent examples of line defects, work as order parameters in characterizing phase structures of gauge theories \cite{Wilson:1974sk, tHooft:1977nqb} while in condensed matter physics spin impurities of electron systems trigger the Kondo effect \cite{Kondo:1964, LudwigAffleck:1991, Affleck:1995ge}.
It is important to understand the universal aspects of extended objects in quantum field theories.
Theories with defects different at microscopic scales tend to fall into the same universality class at macroscopic scales, where the system becomes scale invariant and is typically described by defect conformal field theory (defect CFT, DCFT). 
DCFT aims at exploring critical phenomena in the presence of defects and allows us to access the data characterizing defects through e.g., critical exponents.

One of the simplest models expected to be DCFT at the critical point is the O$(N)$ vector model with a line defect in $d=(4-\epsilon)$ dimensions \cite{Allais:2014fqa,Cuomo:2021kfm}:\footnote{We normalize the kinematic part of the action so that the two-point function of $\Phi_{1}^{\alpha}$ is unit-normalized when $h=0$. This normalization is different from the one used in \cite{Allais:2014fqa,Cuomo:2021kfm}.} 
\begin{align}\label{eq:magnetized model}
    I=\int \d^d x \,\left[ \frac{1}{2\,(d-2)\,\Omega_{d-1}}\,|\partial\Phi_1|^2 + \frac{\lambda\,  \mu^{\epsilon}}{4!}\, |\Phi_1|^4\right] - h\, \mu^{\frac{\epsilon}{2}} \int \d\hat{x}^1\,\Phi^{1}_{1} \ ,
\end{align}
where $\Phi_{1}^\alpha$ $(\alpha=1,\cdots ,N)$ is an O$(N)$ vector field and $\Omega_{d-1}=2\pi^{d/2}/\Gamma(d/2)$ is the volume of a $(d-1)$-sphere. 
The line defect extends parallel to the first axis $\hat{x}^1$ and breaks the O$(N)$ symmetry down to O$(N-1)$.
Hence, we will make a distinction between the first O$(N)$ index $\alpha=1$ and the rest $\alpha=\hat{\alpha}=2,\cdots,N$.
The coordinates of $\mathbb{R}^d$ are also decomposed into the parallel and transverse directions to the defect:
\begin{align}
    x^\mu=(\hat{x}^1,x_\perp^i)\ ,  \qquad i=2,\cdots,d\ .
\end{align}
This model shows critical behaviors when we tune the bulk coupling $\lambda$ and defect coupling $h$ to the values:
\begin{align}\label{eq:critical coupling constants}
    \lambda_\ast=\frac{3}{\pi^2\,(N+8)}\,\epsilon+O(\epsilon^2)\ , \qquad  h_\ast^2=\frac{N+8}{4\pi^2}+O(\epsilon)\ .
\end{align}
We refer to the fixed point with $(\lambda, h) = (\lambda_{\ast}, h_{\ast})$ as the Wilson-Fisher fixed point.
At criticality, the symmetry group of this model enhances to
\begin{align}\label{eq:residual symmetry group on the defect}
    \mathrm{SL}(2,\mathbb{R})\times \mathrm{SO}(d-1)\times \mathrm{O}(N-1)\ ,
\end{align}
where $\mathrm{SL}(2,\mathbb{R})$ is the conformal symmetry parallel to the line defect and $\mathrm{SO}(d-1)$ is the rotational symmetry around the defect.
Defect local operators are classified according to the symmetry group \eqref{eq:residual symmetry group on the defect} and the anomalous dimensions of several defect operators have been calculated by \cite{Cuomo:2021kfm} in perturbation theory.

A textbook approach to critical phenomena is based on renormalization group flow with diagrammatic calculations.
On the other hand, the recent developments of conformal bootstrap have shown that conformal symmetry is strong enough to constrain the dynamics of the theory at fixed points \cite{Rattazzi:2008pe,Poland:2018epd}.
The bootstrap approach has applications to the O$(N)$ vector model in the presence of a boundary \cite{Liendo:2012hy,Bissi:2018mcq,Dey:2020jlc,Padayasi:2021sik,Gliozzi:2015qsa} or defects \cite{Gimenez-Grau:2022ebb,Gimenez-Grau:2022czc,Bianchi:2022sbz}.
More recently, Rychkov and Tan developed an axiomatic method to examine critical behaviors \cite{Rychkov:2015naa} with the aim to bridge between diagrammatic and bootstrap approaches. 
By postulating a set of axioms, they succeeded in reproducing the leading anomalous dimensions of local operators in the $(4-\epsilon)$-dimensional $\mathrm{O}(N)$ model at the Wilson-Fisher fixed point without resorting to perturbation theory. Their axiomatic method has been extended and applied to various models \cite{Basu:2015gpa,Ghosh:2015opa,Raju:2015fza,Nii:2016lpa,Giombi:2017rhm,Yamaguchi:2016pbj,Giombi:2020rmc,Soderberg:2017oaa,Herzog:2022jlx,Dey:2020jlc}.

In this paper, we leverage the Rychkov-Tan method to investigate the O$(N)$ vector model with a line defect \eqref{eq:magnetized model} at criticality.
We assume the bulk coupling is tuned to the value $\lambda_\ast$ at the Wilson-Fisher fixed point and the whole system is at a conformal fixed point with the defect coupling $h$ undetermined.
One of our main results is to show that the critical defect coupling is uniquely fixed by defect conformal symmetry to the value $h_\ast$ at the Wilson-Fisher fixed point \eqref{eq:critical coupling constants}.
We note that the axioms we employ incorporate the equation of motion for bulk operators but do not introduce any direct relation between defect operators.
Nevertheless, by combining the axioms with defect operator expansions (DOEs) we obtain nontrivial identities for defect operators, one of which turns out to yield the critical value $h_\ast$.
We believe this machinery is not limited to the O$(N)$ vector model but is applicable to more general classes of DCFTs.

We also derive the leading anomalous dimensions of various defect local operators including composite ones by the axiomatic approach.
While it is straightforward to derive the dimensions of the lowest-lying operators, the same strategy does not apply to higher-order (composite) operators unless considering the analyticity of correlation functions.
Reassuringly, the resulting dimensions precisely agree with those obtained by diagrammatic calculations in \cite{Cuomo:2021kfm}.
The properties of the defect local operators considered in this paper are summarized in table \ref{tab:list of anomalous dimensions}.

This paper is organized as follows.
In section \ref{sec:Rychkov-Tan Axioms}, we review the axiomatic approach to critical phenomena by Rychkov and Tan, followed by the extension to the case with defects.
In section \ref{eq:Free theory with a line defect}, we proceed to examine the correlation functions and the structures of DOE in the free theory in four dimensions.
In section \ref{1st order: scalar}, we reproduce the critical defect coupling and the leading anomalous dimensions of the lowest-lying defect local operator: $\widehat{W}_{1}^{\alpha}$.
In section \ref{1st order: transverse spin}, we calculate the leading anomalous dimensions of the defect local operators with transverse spin associated with the rotation group SO$(d-1)$ around the line defect: $\widehat{U}_{s}^{\alpha}$.
In section \ref{sec:defect composite operator}, we compute the leading anomalous dimensions of defect composite operators: $\widehat{S}_{\pm}$,  $\widehat{V}^{\hat{\alpha}}$, $\widehat{T}^{\hat{\alpha}\hat{\beta}}$, which are in scalar, vector and tensor representations under the global symmetry O$(N-1)$, respectively. Appendix \ref{app:bdd 3pt} is devoted to the review of the conformal block expansion of the bulk-defect-defect three-point functions in DCFT.

\paragraph{Notes added:} Section \ref{1st order: transverse spin} overlaps with appendix C of a recent paper \cite{Giombi:2022vnz}, where the conformal dimensions of defect local operators with transverse spin were derived in a slightly different manner.

\begin{table}[ht]
\centering
\renewcommand{\arraystretch}{1.7}
\begin{tabular}{cccccc}
\toprule
Operators  & Dimension & SO$(d-1)$ rep. & O$(N-1)$  rep. & Free limit \\ \midrule
$\widehat{W}_1^{\,1}$  & \eqref{conf dim: defect local scalar}  & scalar & singlet &$\widehat{\Phi}^{\,1}_{1} $ \\
$\widehat{W}_1^{\hat{\alpha}}$  & \eqref{conf dim: defect local scalar}  & scalar & vector&$\widehat{\Phi}_{1}^{\hat{\alpha}}$ \\
 $\widehat{U}_{i_1\cdots i_s}^{\,1}$  & \eqref{eq:conf dim:transverse spin}  & tensor & singlet &$\widehat{\Phi}_{s+1,i_1\cdots i_s}^{\,1}$ \\
$\widehat{U}_{i_1\cdots i_s}^{\,\hat{\alpha}}$  & \eqref{eq:conf dim:transverse spin}  & tensor & vector &$\widehat{\Phi}_{s+1,i_1\cdots i_s}^{\,\hat{\alpha}}$\\
$\widehat{S}_\pm$  & \eqref{conf dim: scalars}  & scalar & singlet &$\left\{ |\widehat{\Phi}_{1}^{\, 1}|^2, |\widehat{\Phi}_{1}^{\, \hat{\alpha}}|^2 \right\}$\\
$\widehat{V}^{\hat{\alpha}}$  &\eqref{conf dim: vector} & scalar & vector &$\widehat{\Phi}_{1}^{\,1}\widehat{\Phi}_{1}^{\,\hat{\alpha}}$ \\
$\widehat{T}^{\hat{\alpha}\hat{\beta}}$ & \eqref{conf dim: tensor} & scalar & tensor & $\displaystyle\widehat{\Phi}_{1}^{\,(\hat{\alpha}}\widehat{\Phi}_{1}^{\,\hat{\beta})}$\\
\bottomrule
\end{tabular}
\caption{Summary of the conformal dimensions of defect local operators reproduced in this paper. They are classified by the symmetry group on the line defect \eqref{eq:residual symmetry group on the defect}.
The free limits of $\widehat{S}_\pm$ become linear combinations of $|\widehat{\Phi}_{1}^{\, 1}|^2$ and $|\widehat{\Phi}_{1}^{\, \hat{\alpha}}|^2$.}
\label{tab:list of anomalous dimensions}
\end{table}

\section{Review of Rychkov-Tan method} \label{sec:Rychkov-Tan Axioms}
This section describes Rychkov-Tan's axiomatic approach \cite{Rychkov:2015naa} to critical phenomena.
After a brief review of their framework, we show that their axioms can be generalized to the case with a conformal defect of planer or spherical shape with some modifications.
We note that all the statements below are valid for any dimensional conformal defect, including a boundary.

\subsection{Axioms in CFT}\label{eq:homogeneous CFT axiom}
The first axiom Rychkov-Tan postulates is about the conformal symmetry at the Wilson-Fisher fixed point:
\begin{itemize}\setlength{\leftskip}{8mm}
\begin{shaded}
    \item[\textbf{Axiom \mylabel{cftaxiom1}{\RomanNumeralCaps{1}}.}] The theory at the Wilson-Fisher fixed point has conformal symmetry. 
\end{shaded}
\end{itemize}
It follows that the operator product expansions (OPEs) can be used at the Wilson-Fisher fixed point.
For scalar operators, they are schematically written as
\begin{align}\label{eq:bulk scalar OPE}
    \CO_{\Delta_1}(x)\times\CO_{\Delta_2}(0)
    \supset \, \frac{c_{12k}}{|x|^{\Delta_1+\Delta_2-\Delta_k}}\,\left[1+c_1\,x^\mu\partial_\mu+\cdots \right]\,\CO_{\Delta_k}(0)\ ,
\end{align}
where $c_1$ is a coefficient fixed by conformal symmetry.

The second axiom is about the relation between operators at the free fixed point ($\epsilon=0$) and the Wilson-Fisher fixed point ($\epsilon\neq0$):
\begin{itemize}\setlength{\leftskip}{10mm}
    \begin{shaded}
    \vspace{-3mm}
    \item[\textbf{Axiom \mylabel{cftaxiom2}{\RomanNumeralCaps{2}}.}] For every local operator $\CO_{\text{free}}$ in the free theory ($\epsilon=0$), there exists a local
operator at the Wilson-Fisher fixed point ($\epsilon\neq0$), $\CO_{\text{WF}}$, which tends to $\CO_{\text{free}}$ in the free limit $\epsilon \rightarrow 0$: $\lim_{\epsilon \rightarrow 0}\CO_{\text{WF}}=\CO_{\text{free}}$. 
\end{shaded}
\end{itemize}
Axiom \ref{cftaxiom2} implies that e.g., for free field operators $\Phi_1^\alpha$ and $\Phi_3^\alpha\equiv \Phi_{1}^{\alpha}|\Phi_{1}|^2$ there exist corresponding operators $W_1^\alpha$ and $W_3^\alpha$ at the Wilson-Fisher fixed point. 

Note that both the $(4-\epsilon)$-dimensional free theory and the theory at the Wilson-Fisher point satisfy Axioms \ref{cftaxiom1} and \ref{cftaxiom2}.
Hence, to make a distinction between the two theories we must add the third axiom: 
\begin{itemize}\setlength{\leftskip}{12mm}
\begin{shaded}
    \item[\textbf{Axiom \mylabel{cftaxiom3}{\RomanNumeralCaps{3}}.}]
\vspace{-3mm}
At the Wilson-Fisher fixed point, $W_1^\alpha$ and $W_3^\alpha$ are related by the following equation of motion:
    \begin{align}\label{eq:classical EoM O(N) homogeneous CFT}
    \Box_x \, W_1^\alpha(x)=\kappa\,W_3^\alpha(x)\  ,
\end{align}
 where $\Box_x$ is the Laplacian in $d=(4-\epsilon)$ dimensions. 
 \end{shaded}
\end{itemize}
Two operators $\Phi_{1}^{\alpha}$ and $\Phi_{3}^{\alpha}$ are primary in the free theory and constitute conformal multiplets independently. 
Axiom \ref{cftaxiom3} asserts that at the Wilson-Fisher fixed point $W_3^\alpha$ turns into a descendant of $W_1^\alpha$ and their conformal multiplets recombine.
This multiplet recombination makes the interacting theory different from the free one.

With these axioms, Rychkov and Tan have determined $\kappa$ in \eqref{eq:classical EoM O(N) homogeneous CFT} and reproduced the leading anomalous dimensions as follows \cite{Rychkov:2015naa}. 

First, Axiom \ref{cftaxiom1} allows us to consider the OPE of $W_1^\alpha$:
\begin{align}\label{eq:bulk W1W1 OPE}
    W_1^\alpha(x)\times W_1^\beta(0)\supset \frac{\delta^{\alpha\beta}\,c_1}{|x|^{2\Delta_{W_1}}}\,\bm{1}+\cdots \ ,
\end{align}
with $\bm{1}$ being the identity operator. 
Then, Axiom \ref{cftaxiom2} requires that the free limit of \eqref{eq:bulk W1W1 OPE} should be
\begin{align}
    \Phi_1^\alpha(x)\times \Phi_1^\beta(0)\supset \frac{\delta^{\alpha\beta}}{|x|^2}\,\bm{1}+\cdots \, .
\end{align}
We thus conclude $c_{1}=1+O(\epsilon)$ and $\Delta_{W_1}=1+O(\epsilon)$.

Next, acting the Laplacian on the LHS of \eqref{eq:bulk W1W1 OPE} and using Axiom \ref{cftaxiom3}, we find
\begin{align}\label{eq:bulk W3W3 OPE}
\begin{aligned}
    W_3^\alpha&(x)\times W_3^\beta(0)\\
    &\supset \frac{4\,\Delta_{W_1}(\Delta_{W_1}+1)(2\Delta_{W_1}+2-d)(2\Delta_{W_1}+4-d)}{\kappa^2}\,\frac{\delta^{\alpha\beta}\,c_1}{|x|^{2\Delta_{W_1}+4}}\,\bm{1}+\cdots \ .
\end{aligned}
\end{align}
On the other hand, when $\epsilon=0$, the LHS of \eqref{eq:bulk W3W3 OPE} should reduce to $\Phi_3^\alpha(x)\times \Phi_3^\beta(0)$ (Axiom \ref{cftaxiom2}), which can be calculated by Wick's theorem as
\begin{align}\label{eq:bulk P3P3 OPE}
\Phi_3^\alpha(x)\times \Phi_3^\beta(0)\supset \frac{2\,(N+2)\,\delta^{\alpha\beta}}{|x|^6}\,\bm{1}+\cdots \ .
\end{align}
Comparing \eqref{eq:bulk W3W3 OPE} and \eqref{eq:bulk P3P3 OPE}, we end up with
\begin{align}
    \Delta_{W_1}=\frac{d-2}{2}+\frac{N+2}{16}\kappa^2+(\text{higher order terms in }\epsilon)\ .
\end{align}
Similar considerations for the OPEs of composite operators such as $\Phi_{2p}\times \Phi_{2p+1}^{\alpha}$ and $\Phi_{2p+1}^{\alpha}\times \Phi_{2p+2}$ with $\Phi_{2p}\equiv|\Phi_1|^{2p},\Phi_{2p+1}^{\alpha}\equiv\Phi_1^\alpha|\Phi_1|^{2p}$ give rise to further constraints on the constants $\kappa$ and $\Delta_{W_1}$.
After solving them one successfully reproduces the known diagrammatic results:
\begin{align}
\kappa=\frac{2}{N+8}\,\epsilon+O(\epsilon^2)\ ,\qquad
\Delta_{W_1}=\frac{d-2}{2}+\frac{N+2}{4\,(N+8)^2}\,\epsilon^2 +O(\epsilon^3)\ .
\end{align}

\subsection{Axioms in DCFT}\label{eq:DCFT axiom}
We move on to the modified axioms adapted to studying critical phenomena in the presence of a defect \cite{Yamaguchi:2016pbj}.

First, Axiom \ref{cftaxiom1} is replaced by    \begin{itemize}\setlength{\leftskip}{11mm}
    \begin{shaded}
    \item[\textbf{Axiom \mylabel{dcftaxiom1}{\RomanNumeralCaps{1}'}.}] In the presence of a defect, the theory at the Wilson-Fisher fixed point has the \textbf{defect conformal symmetry}.
    \end{shaded}
\end{itemize}
This axiom implies that the theory is described by DCFT, which allows two types of local operators; bulk and defect local operators.
Axiom \ref{dcftaxiom1} allows us to exploit DOE as an operator identity between bulk and defect local operators.

We postulate that, in taking $\epsilon\to0$, the Wilson-Fisher DCFT should be reduced to the free theory with a defect. Axiom \ref{cftaxiom2} is not modified in essence but should be restated to include bulk and defect local operators.
  \begin{itemize}\setlength{\leftskip}{13mm}
  \begin{shaded}
  \vspace{-3mm}
      \item[\textbf{Axiom \mylabel{dcftaxiom2}{\RomanNumeralCaps{2}'}.}] \textbf{For a bulk/defect local operator} $\CO_{\text{free}}/\widehat{\CO}_{\text{free}}$ in the free theory with a defect, there exists a local operator $\CO_{\text{WF}}/\widehat{\CO}_{\text{WF}}$ at the Wilson-Fisher fixed point, which tends to $\CO_{\text{free}}/\widehat{\CO}_{\text{free}}$ in the limit $\epsilon \to 0$.
  \end{shaded}
\end{itemize}

Axiom \ref{cftaxiom3} holds as it stands without introducing any relation for defect operators:
\begin{itemize}\setlength{\leftskip}{15mm}
\begin{shaded}
    \item[\textbf{Axiom \mylabel{dcftaxiom3}{\RomanNumeralCaps{3}'}.}] At the Wilson-Fisher fixed point, \textbf{two bulk operators} $W_1^\alpha$ and $W_3^\alpha$ are related by the following equation of motion:
    \begin{align}\label{eq:classical EoM O(N)}
    \Box_x \, W_1^\alpha(x)=\kappa\,W_3^\alpha(x)\  ,
\end{align}
where $\Box_x$ is the Laplacian in $d=(4-\epsilon)$ dimensions.
\end{shaded}
\end{itemize}
Notice that we can use the OPEs for bulk operators \eqref{eq:bulk scalar OPE} in DCFT, hence by repeating the same discussion as in the previous section we find
\begin{align}\label{eq:input from bulk}
\kappa=\frac{2}{N+8}\,\epsilon +O(\epsilon^2)\ ,\qquad \Delta_{W_1}=\frac{d-2}{2}+O(\epsilon^2)\ .
\end{align}

\subsection{Structures of DCFT}\label{eq:Structures of DCFT}
We list our notations and give a brief review on DCFT with a $p$-dimensional defect in $d$ dimensions \cite{Billo:2016cpy,Gadde:2016fbj}.

On the $d$-dimensional flat spacetime $\mathbb{R}^d$, we place a $p$-dimensional planer defect at $x^\mu=0$ with $\mu=p+1,\cdots,d$.
We express a bulk point by $x^\mu$ and decompose its coordinate into parallel and transverse directions to the defect:
\begin{align}
    x^\mu=(\hat{x}^a,x_\perp^i)\ ,\qquad a=1,\cdots, p\ ,\qquad i=p+1,\cdots, d\ ,
\end{align}
while we use $\hat{y}$ to denote a point on the defect;
\begin{align}
    \hat{y}^\mu=(\hat{y}^a,0)\ .
\end{align}
 We classify the bulk local operators according to the representations of the full conformal group $\mathrm{SO}(1,d+1)$ and, throughout this paper, focus only on the scalar ones $\CO_{\Delta}$ characterized by the conformal dimension $\Delta$. On the other hand, the defect local operators are labeled by the defect conformal group that is a direct product of the conformal group parallel to the defect $\mathrm{SO}(1,p+1)$ and the rotation around the defect $\mathrm{SO}(d-p)$. We are particularly interested in the scalar defect local operators labeled by conformal dimension $\widehat{\Delta}$: $\widehat{\CO}_{\widehat{\Delta}}$, and the ones carrying transverse spin $s$: $\widehat{\CO}^s_{\widehat{\Delta},i_1\cdots i_s}$. 

The correlation functions of bulk and defect scalars are fixed as
\begin{align}\label{eq:scalar correlator DCFT 2 and 3}
\begin{aligned}
        \langle\,\CO_{\Delta}(x)\,\widehat{\CO}_{\widehat{\Delta}}(\hat{y})\,\rangle
        &=
            \frac{b(\CO,\widehat{\CO})}{|x-\hat{y}|^{2\widehat{\Delta}}\,|x_\perp|^{\Delta-\widehat{\Delta}}}\ ,\\
    \langle\,\widehat{\CO}_{\widehat{\Delta}_1}(\hat{y}_1)\,\widehat{\CO}_{\widehat{\Delta}_2}(\hat{y}_2) \,\widehat{\CO}_{\widehat{\Delta}_3}(\hat{y}_3) \,\rangle
        &=
            \frac{c(\widehat{\CO}_1,\widehat{\CO}_2,\widehat{\CO}_2)}{|\hat{y}_{12}|^{\widehat{\Delta}^+_{12}-\widehat{\Delta}_3}\,|\hat{y}_{23}|^{\widehat{\Delta}^+_{23}-\widehat{\Delta}_1}\,|\hat{y}_{13}|^{\widehat{\Delta}^+_{13}-\widehat{\Delta}_2}}\ , \\
        \langle\,\widehat{\CO}_{\widehat{\Delta}}(\hat{y}_1)\,\widehat{\CO}_{\widehat{\Delta}}(\hat{y}_2)\,\rangle
            &=
                \frac{c(\widehat{\CO},\widehat{\CO})}{|\hat{y}_{12}|^{2\widehat{\Delta}}}\ ,
\end{aligned}
\end{align}
where we used the following shorthand notations:
\begin{align}
    \widehat{\Delta}^\pm_{ij}\equiv \widehat{\Delta}_{i}\pm\widehat{\Delta}_{j}\ ,\qquad \hat{y}_{12}\equiv \hat{y}_1-\hat{y}_2 \ .
\end{align}
The two-point functions involving defect local operators with transverse spin are
\begin{align}\label{eq:spinning correlator DCFT 2 and 3}
    \begin{aligned}
      \langle\,\CO_{\Delta}(x)\,\widehat{\CO}_{\widehat{\Delta}}^{s,i_1\cdots i_s}(\hat{y})\,\rangle
        &=
            \frac{x_\perp^{(i_1}\cdots x_\perp^{i_s)} b(\CO,\widehat{\CO}^s)}{|x-\hat{y}|^{2\widehat{\Delta}}\,|x_\perp|^{\Delta-\widehat{\Delta}+s}}\ ,\\
           \langle\,\widehat{\CO}_{\widehat{\Delta}}^{s,i_1\cdots i_s}(\hat{y}_1)\,\widehat{\CO}^s_{\widehat{\Delta},j_1\cdots j_s}(\hat{y}_2)\,\rangle
            &=
                \frac{\delta_{j_1}^{(i_1}\cdots \delta_{j_s}^{i_s)}\,c(\widehat{\CO}^s,\widehat{\CO}^s)}{|\hat{y}_{12}|^{2\widehat{\Delta}}}\ .
    \end{aligned}
\end{align}
Throughout this paper, the parenthesis appearing in the indices stands for the symmetric and traceless structure:
\begin{align}
    x_\perp^{(i}x_\perp^{j)}=x_\perp^i x_\perp^j-\frac{\delta^{ij}}{d-p}\,|x_\perp|^2\ .
\end{align}

A bulk local operator can be expanded in terms of defect local ones:
\begin{align}\label{eq:btd OPE schematic}
    \CO_{\Delta}(x)
    \supset 
    \sum_{\widehat{\CO}^s} \,\frac{b(\CO,\widehat{\CO}^s)/c(\widehat{\CO}^s,\widehat{\CO}^s)}{|x_\perp|^{\Delta-\widehat{\Delta}+s}}\,x_\perp^{(i_1}\cdots x_\perp^{i_s)}\,\widehat{\CO}^s_{\widehat{\Delta},i_1\cdots i_s}(\hat{x}) \ .
\end{align}
The symbol $\supset$ stands for the DOE, and we do not bother to write descendant terms.
This expansion is fixed for all orders in $|x_\perp|$ to be compatible with \eqref{eq:spinning correlator DCFT 2 and 3} (see \eqref{eq:scalar btd OPE full} for the full form for the scalar case).

We also give a quick review of bulk-defect-defect three-point functions $\langle\, \CO_{\Delta}\,\widehat{\CO}_{\widehat{\Delta}_1}\,\widehat{\CO}_{\widehat{\Delta}_2} \,\rangle$ that are of great importance in section \ref{sec:defect composite operator}, whose details are relegated to appendix \ref{app:bdd 3pt}.
 For simplicity we place $\widehat{\CO}_{\widehat{\Delta}_1}$ and $\widehat{\CO}_{\widehat{\Delta}_2}$ at the origin and infinity by using conformal symmetry.\footnote{The defect local operator at the infinity is defined by $\widehat{\CO}_{\widehat{\Delta}}(\infty)=\lim_{|\hat{y}|\to\infty}\,|\hat{y}|^{2\widehat{\Delta}}\,\widehat{\CO}_{\widehat{\Delta}}(\hat{y})$.}
The bulk-defect-defect three-point function depends on the cross-ratio $\upsilon=|x_\perp|^2/|x|^2$ and has the following conformal block expansion:
\begin{align}\label{eq:conformal block expansion main}
    \begin{aligned}
              \langle\, \CO_{\Delta}(x)\,\widehat{\CO}_{\widehat{\Delta}_1}(0)\,\widehat{\CO}_{\widehat{\Delta}_2}(\infty) \,\rangle&=
                    \frac{1}{|x_\perp|^{\Delta} \,|x|^{\widehat{\Delta}^-_{12}}} \,\sum_{\widehat{\CO}}\, \frac{b(\CO,\widehat{\CO})\,c(\widehat{\CO},\widehat{\CO}_1,\widehat{\CO}_2)}{c(\widehat{\CO},\widehat{\CO})}\,G^{\widehat{\Delta}^-_{12}}_{\widehat{\Delta}}\left(\frac{|x_\perp|^2}{|x|^2}\right)\ ,\\
       G^{\widehat{\Delta}^-_{12}}_{\widehat{\Delta}}(\upsilon)&=\upsilon^{\widehat{\Delta}/2}\,{}_2F_1\left(\frac{\widehat{\Delta}+\widehat{\Delta}^-_{12}}{2},\frac{\widehat{\Delta}-\widehat{\Delta}^-_{12}}{2};\widehat{\Delta}+1-\frac{p}{2};\upsilon\right)\ .             
    \end{aligned}
\end{align}
Note that only scalar operators contribute to the conformal block expansion in \eqref{eq:conformal block expansion main} as the defect three-point functions of two scalars and one spinning operator vanish.

\section{The free O$(N)$ model in four dimensions with a line defect}\label{eq:Free theory with a line defect}
In this section, we investigate the free O$(N)$ model ($\lambda=0$) in four dimensions with the line defect defined by \eqref{eq:magnetized model}.
We focus on the correlation functions with the line defect and DOEs associated with the bulk operators $\Phi_1^\alpha$ and $\Phi_3^\alpha$, which are necessary for the later section.

\subsection{Correlation functions}\label{sec:Correlation functions in free theory}
Let $\langle \,\cdots\,\rangle_0$ be a correlation function in the free theory with no defect coupling:\footnote{The path integral measure is normalized such that the expectation value of the identity operator is one: $\langle\,{\bf 1}\,\rangle_0 = 1$.}
\begin{align}
    \langle \,\cdots\,\rangle_0
        \equiv
        \int \CD \Phi_1\, (\cdots)\,\exp\left( -\frac{1}{8\pi^2}\int\d^4 x\,|\partial \Phi_1|^2\right) \ .
\end{align}
Then, we define correlation functions in the presence of the defect by \cite{Kapustin:2005py, Billo:2016cpy}
\begin{align}
    \langle\,\cdots\,\rangle \equiv \frac{\langle\,\cdots \,e^{h\,\int \d \hat{y}^1\,\Phi_1^1}\,\rangle_0}{\langle\,e^{h\,\int \d \hat{y}^1\,\Phi_1^1}\,\rangle_0}\ .
\end{align}

We begin by considering the two-point function without a defect, which satisfies the following differential equation:
\begin{align}\label{eq:4d free scalar propagator EoM}
    \Box_{x_1}\,\langle\,\Phi_{1}^\alpha(x_1)\,\Phi_{1}^\beta(x_2)\,\rangle_0
        =
            4\pi^2\,\delta^{\alpha\beta}\,\delta^d(x_1-x_2)\ .
\end{align}
The solution to this equation is given by
\begin{align}\label{eq:free scalar propargator}
  \langle\,\Phi_1^\alpha(x_1)\,\Phi_1^\beta(x_2)\,\rangle_0=\frac{\delta^{\alpha\beta}}{|x_1-x_2|^{2}}\ .
\end{align}
Once the defect coupling is turned on, the one-point function of $\Phi_1^\alpha$ no longer vanishes \cite[section 5.4]{Billo:2016cpy}:
\begin{align}\label{eq:bulk one point phi}
              \langle\,\Phi_1^\alpha(x)\,\rangle
             =
                \frac{\delta^{\alpha1}\,\hat{h}}{|x_\perp|}\ ,
\end{align}
where we define $\hat{h}$ by
\begin{align}
    \hat{h}\equiv\pi\,h\ .
\end{align}
In what follows, we use $\hat{h}$ instead of the defect coupling $h$ for convenience.
The bulk two-point function can be calculated similarly:
\begin{align}\label{eq:bulk two point phi}
              \langle\,\Phi_1^\alpha(x_1)\,\Phi_1^\beta(x_2)\,\rangle=\frac{\delta^{\alpha1}\,\delta^{\beta1}\,\hat{h}^2}{|x_{1,\perp}|\,|x_{2,\perp}|}+\frac{\delta^{\alpha\beta}}{|x_1-x_2|^{2}}\ .
\end{align}
 We now define a defect local operator $\widehat{\Phi}_1^{\,\alpha}$ by
 \begin{align}
     \widehat{\Phi}_1^{\,\alpha}(\hat{x})\equiv \lim_{|x_\perp|\to0}\,\Phi_1^\alpha(x)\ .
 \end{align}
Two-point functions involving $\widehat{\Phi}_1^{\,\alpha}$ can be deduced from \eqref{eq:bulk two point phi}:
\begin{align}\label{eq:fund 2pt}
    \langle\,\Phi_1^\alpha(x)\,\widehat{\Phi}_1^{\,\beta}(\hat{y})\,\rangle=\frac{\delta^{\alpha\beta}}{|x-\hat{y}|^2} \ ,\qquad
        \langle\,\widehat{\Phi}_1^{\,\alpha}(\hat{y}_1)\,\widehat{\Phi}_1^{\,\beta}(\hat{y}_2)\,\rangle=\frac{\delta^{\alpha\beta}}{|\hat{y}_{12}|^{2}} \ .
\end{align}
We are also interested in the defect local operators with transverse spin:
\begin{align}
    \widehat{\Phi}_{s+1}^{\,\alpha,i_1\cdots i_s}(\hat{x})\equiv \lim_{|x_\perp|\to0}\,\partial^{(i_1}\cdots\partial^{i_s)}\Phi_1^{\alpha}(x)\ .
\end{align}
Some of their correlation functions are given by\footnote{We have to use the regularization scheme that respects defect conformal invariance. To be more specific, we start with correlators consisting only of bulk fields, act derivatives;
\begin{align*}
\begin{aligned}     \langle\,\partial^{(i_1}\cdots\partial^{i_s)}\Phi_1^\alpha(x_1)\,\Phi_1^\beta(x_2)\,\rangle&=\frac{(-2)^s\,s!\,\delta^{\alpha\beta}\,x_{\perp,12}^{(i_1}\cdots x_{\perp,12}^{i_s)}}{|x_{12}|^{2(s+1)}}+(\text{singular terms in $|x_{1,\perp}|$})\ ,
\end{aligned}
\end{align*}
and then take $|x_{1,\perp}|\to 0$ limit, dropping off singular terms with negative powers of $|x_{1,\perp}|$, to obtain \eqref{eq:transverse spin free 1}. Similar manipulations for \eqref{eq:transverse spin free 1} lead to \eqref{eq:transverse spin free 2}.}
 \begin{align}
               \langle\,\widehat{\Phi}_{s+1}^{\,\alpha,i_1\cdots i_s}(\hat{y})\,\Phi_1^\beta(x)\,\rangle&=2^s\,s!\,\delta^{\alpha\beta}\,\frac{x_{\perp}^{(i_1}\cdots x_{\perp}^{i_s)}}{|x-\hat{y}|^{2(s+1)}}\ ,\label{eq:transverse spin free 1}\\
               \langle\,\widehat{\Phi}_{s+1}^{\,\alpha,i_1\cdots i_s}(\hat{y}_1)\,\widehat{\Phi}_{s+1}^{\,\beta,j_1\cdots j_s}(\hat{y}_2)\,\rangle&=2^s\,(s!)^2\,\delta^{\alpha\beta}\,\frac{\delta^{(i_1}_{j_1}\cdots \delta^{i_s)}_{j_s}}{|\hat{y}_{12}|^{2(s+1)}}\ .\label{eq:transverse spin free 2}
\end{align}
Using the result obtained so far, any correlation functions can be computed via Wick's theorem. 
In the following, we enumerate the results relevant to this paper. 
\paragraph{One- and two-point functions.} 
We define $\widehat{\Phi}_3^{\,\alpha}(\hat{x})$ by
\begin{align}
    \widehat{\Phi}_3^{\,\alpha}(\hat{x})\equiv\lim_{|x_\perp|\to0}\,\Phi_3^\alpha(x)\ .
\end{align}
Then, we obtain the correlation functions below:
\begin{align}
    \begin{aligned}
       \langle\,\Phi_3^\alpha(x)\,\rangle&=\frac{\delta^{\alpha1}\,\hat{h}^3}{|x_\perp|^3}\ ,&\qquad
          \langle\,\Phi_3^\alpha(x)\,\widehat{\Phi}_1^{\,\beta}(\hat{y})\,\rangle&= \frac{\hat{h}^2\,(1 + 2\,\delta^{\alpha1})\,\delta^{\alpha\beta}}{|x-\hat{y}|^2\,|x_\perp|^2} \ ,\\
           \langle\,\Phi_3^\alpha(x)\,\widehat{\Phi}_3^{\,\beta}(\hat{y})\,\rangle&= \frac{2\,(N+2)\,\delta^{\alpha\beta}}{|x-\hat{y}|^6}\ ,&\qquad
        \langle\,\widehat{\Phi}_3^{\,\alpha}(\hat{y}_1)\,\widehat{\Phi}_3^{\,\beta}(\hat{y}_2)\,\rangle&= \frac{2\,(N+2)\,\delta^{\alpha\beta}}{|\hat{y}_{12}|^6}\ ,
    \end{aligned}
\end{align}
\begin{align}
          \langle\,\widehat{\Phi}_{s+1}^{\,\alpha,i_1\cdots i_s}(\hat{y})\,\Phi_3^\beta(x)\,\rangle&=\hat{h}^2\,(1 + 2\,\delta^{\alpha1})\,2^s\,s!\,\delta^{\alpha\beta}\,\frac{x_{\perp}^{(i_1}\cdots x_{\perp}^{i_s)}}{|x-\hat{y}|^{2(s+1)}}\ .
\end{align}

\paragraph{Three-point functions.}\label{app:Three-point functions}
We consider the following three-point functions:
\begin{align}
    \langle\,  \Phi_1^{\alpha}(x)\, \widehat{\Phi}_1^{\,\beta}(\hat{y}_1)\,\widehat{\Phi}_2(\hat{y}_2) \,\rangle\ ,\qquad     \langle\,  \Phi_3^{\alpha}(x)\, \widehat{\Phi}_1^{\,\beta}(\hat{y}_1)\,\widehat{\Phi}_2(\hat{y}_2) \,\rangle \ ,
\end{align}
with $\widehat{\Phi}_2$ being the following composite defect local operators:
\begin{align}\label{eq:second order defect local app}
\widehat{\Phi}_2 \in \left\{|\widehat{\Phi}_1^{\,1}|^{2}\ ,~ |\widehat{\Phi}_{1}^{\,\hat{\gamma}}|^{2}\ ,~ \widehat{\Phi}_{1}^{\,1}\widehat{\Phi}_{1}^{\,\hat{\gamma}}\ , ~ \widehat{\Phi}_{1}^{\,(\hat{\gamma}}\widehat{\Phi}_{1}^{\,\hat{\sigma})}\right\}\ .
\end{align}
The first two operators are O$(N-1)$ scalars, whereas the third one is an O$(N-1)$ vector. The last operator $\widehat{\Phi}_1^{\,(\hat{\gamma}}\widehat{\Phi}_1^{\,\hat{\sigma})}$ is a rank two O$(N-1)$ symmetric traceless tensor:
\begin{align}
\begin{aligned}
        |\widehat{\Phi}_{1}^{\,1}|^{2}(\hat{x})&\equiv\lim_{|x_\perp|\to0}\,\Phi_1^{1}\Phi_1^{1}(x)\ ,&\qquad |\widehat{\Phi}_{1}^{\,\hat{\gamma}}|^{2}(\hat{x})&\equiv\lim_{|x_\perp|\to0}\,\sum_{\hat{\gamma}=2}^N\,\Phi_1^{\,\hat{\gamma}}\Phi_1^{\,\hat{\gamma}}(x)\ ,\\
        \widehat{\Phi}_1^{\,1}\widehat{\Phi}_1^{\,\hat{\gamma}}(\hat{x})&\equiv\lim_{|x_\perp|\to0}\,\Phi_1^1\Phi_1^{\hat{\gamma}}(x)\ ,&\qquad     \widehat{\Phi}_1^{\,(\hat{\gamma}}\widehat{\Phi}_1^{\,\hat{\sigma})}(\hat{x})&\equiv \lim_{|x_\perp|\to0}\,\Phi_1^{\,(\hat{\gamma}}\Phi_1^{\,\hat{\sigma})}(x) \ .
\end{aligned}
\end{align}
For any operators listed in \eqref{eq:second order defect local app}, the three-point functions take similar forms:
 \begin{align}
      \langle\,  \Phi_1^{\alpha}(x)\, \widehat{\Phi}_1^{\,\beta}(\hat{y}_1)\,\widehat{\Phi}_2(\hat{y}_2) \,\rangle&=\frac{c(\widehat{\Phi}_1^{\alpha},\widehat{\Phi}_1^{\,\beta},\widehat{\Phi}_2)}{|x-\hat{y}_2|^2\,|\hat{y}_{12}|^2}\ ,\\
      \langle\,  \Phi_3^{\alpha}(x)\, \widehat{\Phi}_1^{\,\beta}(\hat{y}_1)\,\widehat{\Phi}_2(\hat{y}_2) \,\rangle&=\frac{(1+2\,\delta^{\alpha1})\,\hat{h}^2\,c(\widehat{\Phi}_3^{\alpha},\widehat{\Phi}_1^{\,\beta},\widehat{\Phi}_2)}{|x-\hat{y}_2|^2\,|\hat{y}_{12}|^2\,|x_\perp|^2}+ \frac{c(\Phi_3^{\alpha},\widehat{\Phi}_1^{\,\beta},\widehat{\Phi}_2)}{|x-\hat{y}_1|^2\,|x-\hat{y}_2|^4} \ .\label{eq:bdd three point Phi2}
 \end{align}
Here, $c(\widehat{\Phi}_1^{\alpha},\widehat{\Phi}_1^{\,\beta},\widehat{\Phi}_2)$ and $c(\widehat{\Phi}_3^{\alpha},\widehat{\Phi}_1^{\,\beta},\widehat{\Phi}_2)$ are the defect three-point coefficients listed in table \ref{tab:list three-point coeff}:
\begin{align}\label{eq:defect three-point phi2}
\begin{aligned}
       \langle\,\widehat{\Phi}_1^{\alpha}(\hat{x})\,\widehat{\Phi}_1^{\,\beta}(\hat{y}_1)\,\widehat{\Phi}_2(\hat{y_2})\,\rangle&= \frac{c(\widehat{\Phi}_1^{\alpha},\widehat{\Phi}_1^{\,\beta},\widehat{\Phi}_2)}{|\hat{x}-\hat{y}_2|\,|\hat{y}_{12}|^2}\ ,\\
   \langle\,\widehat{\Phi}_3^{\alpha}(\hat{x})\,\widehat{\Phi}_1^{\,\beta}(\hat{y}_1)\,\widehat{\Phi}_2(\hat{y_2})\,\rangle&= \frac{c(\widehat{\Phi}_3^{\alpha},\widehat{\Phi}_1^{\,\beta},\widehat{\Phi}_2)}{|\hat{x}-\hat{y}_1|^2\,|\hat{x}-\hat{y}_2|^4}\ .
\end{aligned}
\end{align}

\begin{table}[t]
\centering
\renewcommand{\arraystretch}{1.5}
\begin{tabular}{>{\centering}m{2.2cm}>{\centering}m{4cm}>{\centering\arraybackslash}m{4cm}}
\toprule
$\widehat{\Phi}_2$  & $c(\widehat{\Phi}_1^{\alpha},\widehat{\Phi}_1^{\,\beta},\widehat{\Phi}_2)$ & $c(\widehat{\Phi}_3^{\alpha},\widehat{\Phi}_1^{\,\beta},\widehat{\Phi}_2)$ \\ \midrule
$|\widehat{\Phi}^{\,1}|^{2}$  & $2\,\delta^{\alpha1}\delta^{\beta1}$  & $2\,\delta^{\alpha\beta}\,+4\,\delta^{\alpha1}\delta^{\beta1}$ \\
$|\widehat{\Phi}_{1}^{\, \hat{\alpha}}|^2$  & $2\,\delta^{\alpha\beta}-2\,\delta^{\alpha1}\delta^{\beta1}$  & $2\,(N+1)\,\delta^{\alpha\beta}-4\,\delta^{\alpha1}\delta^{\beta1}$  \\
 $\widehat{\Phi}^{\,1}\widehat{\Phi}^{\,\hat{\gamma}}$  &$\delta^{\alpha1}\delta^{\beta\hat{\gamma}}+\delta^{\beta1}\delta^{\alpha\hat{\gamma}}$  & $\,\delta^{\alpha1}\delta^{\beta\hat{\gamma}}+2\,\delta^{\beta1}\delta^{\alpha\hat{\gamma}}$  \\
$\widehat{\Phi}^{\,(\hat{\gamma}}\widehat{\Phi}^{\,\hat{\sigma})}$  & $2\,\delta^{\alpha(\hat{\gamma}}\delta^{\hat{\sigma})\beta}$  & $4\,\delta^{\alpha(\hat{\gamma}}\delta^{\hat{\sigma})\beta}$ \\
\bottomrule
\end{tabular}
\caption{List of defect three-point coefficients in \eqref{eq:defect three-point phi2}.
}
\label{tab:list three-point coeff}
\end{table}

\paragraph{Three-point functions for $N=1$.}\label{app:Three-point functions N=1}
When $N=1$ (section \ref{sec:N=1 anomalous dimension}), the defect three-point functions can be calculated as follows:
\begin{align}
\langle\, \widehat{\Phi}_1(\hat{x})\,\widehat{\Phi}_p(\hat{y}_1) \,\widehat{\Phi}_{p+1}(\hat{y}_2)\,\rangle&=\frac{c(\widehat{\Phi}_1,\widehat{\Phi}_p,\widehat{\Phi}_{p+1})}{|\hat{x}-\hat{y}_2|^2\,|\hat{y}_{12}|^{2p}}\ ,\\
\langle\, \widehat{\Phi}_3(\hat{x})\,\widehat{\Phi}_p(\hat{y}_1) \,\widehat{\Phi}_{p+1}(\hat{y}_2)\,\rangle&=\frac{c(\widehat{\Phi}_1,\widehat{\Phi}_p,\widehat{\Phi}_{p+1})}{|\hat{x}-\hat{y}_2|^2\,|\hat{x}-\hat{y}_2|^4\,|\hat{y}_{12}|^{2p-2}}\ ,
\end{align}
where
\begin{align}\label{eq: def of p-order compodite op}
    \widehat{\Phi}_{p}(\hat{x})\equiv \lim_{|x_\perp|\to0} \,|\Phi_{1}| ^p(x) \ ,
\end{align}
and 
\begin{align}
    c(\widehat{\Phi}_1,\widehat{\Phi}_p,\widehat{\Phi}_{p+1})=(p+1)!\ ,\qquad c(\widehat{\Phi}_3,\widehat{\Phi}_p,\widehat{\Phi}_{p+1})=3p\,(p+1)!\ .
\end{align}
The bulk-defect-defect three-point functions are
\begin{align}
\langle\, \Phi_1(x)\,\widehat{\Phi}_p(\hat{y}_1) \,\widehat{\Phi}_{p+1}(\hat{y}_2)\,\rangle&=\frac{c(\widehat{\Phi}_1,\widehat{\Phi}_p,\widehat{\Phi}_{p+1})}{|x-y_2|^2\,|\hat{y}_{12}|^{2p}}\ ,\\
\langle\, \Phi_3(x)\,\widehat{\Phi}_p(\hat{y}_1) \,\widehat{\Phi}_{p+1}(\hat{y}_2)\,\rangle&=\frac{3\,\hat{h}\,c(\widehat{\Phi}_1,\widehat{\Phi}_p,\widehat{\Phi}_{p+1})}{|x-y_2|^2\,|\hat{y}_{12}|^{2p}\,|x_\perp|^2}+\frac{c(\widehat{\Phi}_1,\widehat{\Phi}_p,\widehat{\Phi}_{p+1})}{|x-y_2|^2\,|x-y_2|^4\,|\hat{y}_{12}|^{2p-2}}\ .
\end{align}

\subsection{Defect operator expansions}\label{eq:Bulk-to-defect operator product expansions free}
We proceed to spell out the DOEs of two bulk local operators $\Phi_1^\alpha$ and $\Phi_3^\alpha$. They are determined by comparing the general form of DOEs \eqref{eq:btd OPE schematic} with the correlation functions derived in the last subsection.

\subsubsection{Defect operator expansion of $\Phi_1^\alpha$}
The DOE of $\Phi_1^\alpha$ turns out to be
\begin{align}\label{eq:phi1 OPE Wilsonline}
         \Phi_1^\alpha(x)&=\frac{\delta^{\alpha1}\,\hat{h}}{|x_\perp|} \,\bm{1}+\widehat{\Phi}_1^{\alpha}(\hat{x})+\sum_{s=1}^{\infty}\, \frac{1}{s!}\,x_\perp^{(i_1}\cdots x_\perp^{i_s)}\,\widehat{\Phi}_{s+1,i_1\cdots i_s}^{\,\alpha}(\hat{x}) +(\text{descendants})\ .
\end{align}
One can show that the other defect local operators are absent in this DOE \cite[appendix B.1.1]{Billo:2016cpy}.
To see this, let $\widehat{\CO}_{\widehat{\Delta}_s,i_1\cdots i_s}^{\alpha}(\hat{y})$ be a defect local operator with conformal dimension $\widehat{\Delta}_s$ and $s$ symmetric traceless indices of $\mathrm{SO}(3)$. From the defect conformal symmetry, we have
\begin{align}\label{eq:free btd OPE}
    \Phi_1^\alpha(x)\supset A\,\frac{x_\perp^{(i_1}\cdots x_\perp^{i_s)}}{|x_\perp|^{s+1-\widehat{\Delta}_s}}\,\widehat{\CO}_{\widehat{\Delta},i_1\cdots i_s}^{\alpha}(\hat{x})\ ,
\end{align}
with $A$ being some nonzero constant.
 When we act the Laplacian $\Box_x$ on the LHS of \eqref{eq:free btd OPE}, it should vanish due to the Klein-Gordon equation:
 \begin{align}\label{eq:free btd OPE}
  \Box_x\,\Phi_1^\alpha(x)=0\ .
\end{align}
 On the other hand, the RHS becomes
 \begin{align}\label{eq:RHS of free btd OPE}
     \Box_x\,A\,\frac{x_\perp^{(i_1}\cdots x_\perp^{i_s)}}{|x_\perp|^{s+1-\widehat{\Delta}_s}}\,&\widehat{\CO}_{\widehat{\Delta},i_1\cdots i_s}^{\alpha}(\hat{x})=A\,(\widehat{\Delta}_s-s-1)(\widehat{\Delta}_s+s)\,\frac{x_\perp^{(i_1}\cdots x_\perp^{i_s)}}{|x_\perp|^{s+3-\widehat{\Delta}_s}}\,\widehat{\CO}_{\widehat{\Delta},i_1\cdots i_s}^{\alpha}(\hat{x})\ ,
 \end{align}
 where we used the following identity for the $d$-dimensional Laplacian $\Box_x$ in the presence of a $p$-dimensional planar defect:
\begin{align}\label{eq:action of Lap on spinning btd OPE}
       \Box_x\,\frac{x_\perp^{(i_1}\cdots x_\perp^{i_s)}}{|x_\perp|^\delta}=\delta\,(\delta+2-d+p-2s)\,\frac{x_\perp^{(i_1}\cdots x_\perp^{i_s)}}{|x_\perp|^{\delta+2}}\ .
\end{align}
For \eqref{eq:RHS of free btd OPE} to be zero for a finite $A$, we should have $\widehat{\Delta}_s=s+1$ or $\widehat{\Delta}_s=-s$. The former corresponds to $\widehat{\Phi}_1^{\,\alpha}(\hat{x})$ for $s=0$ and $\widehat{\Phi}_{s+1,i_1\cdots i_s}^{\,\alpha}(\hat{x})$ for $s\geq 1$ as anticipated. On the other hand, the latter case $\widehat{\Delta}_s=-s$ is below the unitarity bound of a one-dimensional CFT on the defect except for the identity operator $\bm{1}$ with $s=0$.

\subsubsection{Defect operator expansion of $\Phi_3^\alpha$}
We find that an infinite number of operators contribute to the DOE of $\Phi_3^\alpha$:
\begin{align}\label{eq:phi3 OPE Wilsonline}
\begin{aligned}
                 \Phi_3^\alpha(x)&\supset\frac{\delta^{\alpha1}\,\hat{h}^3}{|x_\perp|^3} \,\bm{1}+\frac{(1+2\,\delta^{\alpha1})\,\hat{h}^2}{|x_\perp |^{2}}\,\widehat{\Phi}_1^{\,\alpha}(\hat{x})\\
      &\qquad+\frac{(1+2\,\delta^{\alpha1})\,\hat{h}^2}{|x_\perp|^2}\,\sum_{s=1}^{\infty}\, \frac{1}{s!}\,x_\perp^{(i_1}\cdots x_\perp^{i_s)}\,\widehat{\Phi}_{s+1,i_1\cdots i_s}^{\,\alpha}(\hat{x})\\
      &\qquad\qquad+\sum_{n=0}^{\infty}\,\frac{b(\Phi_3^\alpha,\widehat{\mathsf{O}}_{2n+3}^{\,\alpha})}{c(\widehat{\mathsf{O}}_{2n+3}^{\,\alpha},\widehat{\mathsf{O}}_{2n+3}^{\,\alpha})}\,|x_\perp|^{2n}\,\widehat{\mathsf{O}}_{2n+3}^{\,\alpha}(\hat{x})\ . 
\end{aligned}
\end{align}
Here, $\widehat{\mathsf{O}}_{3}^\alpha$ can be identified with $\widehat{\Phi}_3^{\,\alpha}$ and the coefficients in the last line are subject to the relations:
\begin{align}\label{eq:btd 2n+3 OPE coeff}
\frac{b(\Phi^{\alpha}_3,\widehat{\mathsf{O}}_{2n+3}^{\,\alpha})\,c(\widehat{\mathsf{O}}_{2n+3}^{\,\alpha},\widehat{\Phi}_1^{\,\beta},\widehat{\Phi}_2)}{c(\widehat{\mathsf{O}}_{2n+3}^{\,\alpha},\widehat{\mathsf{O}}_{2n+3}^{\,\alpha})}= c(\Phi_3^{\alpha},\widehat{\Phi}_1^{\,\beta},\widehat{\Phi}_2)\, \frac{(-1)^n\,(2)_n}{(n+5/2)_n} \ .
\end{align}
To see how this equality is obtained, let us perform the conformal block expansion of the bulk-defect-defect three-point function \eqref{eq:bdd three point Phi2} and denote the intermediate operators with odd conformal dimensions $2n+3$ $(n=0,1,\cdots)$ by $\widehat{\mathsf{O}}_{2n+3}^{\,\alpha}$:\footnote{We have used $G^{-1}_{1}(\upsilon)=\upsilon^{1/2}$, $G_{2n+3}^{-1}(\upsilon)=\upsilon^{n+3/2}\,{}_2F_1(1+n,2+n;7/2+2n;\upsilon)$ and the following hypergeometric identity \cite[equation (9.1.32)]{luke1969special}; $1=\sum_{n=0}^{\infty}\,\frac{(-1)^n\,(\alpha)_n(\beta)_n}{(n+\lambda)_n\,n!}\,z^n\,{}_2F_1(\alpha+n,\beta+n;\lambda+1+2n;z)$.}
\begin{align}\label{eq:conformal block Phi3}
    \begin{aligned}
       \langle\,\Phi^{\alpha}_3(x)\,\widehat{\Phi}_1^{\,\beta}(0)\,\widehat{\Phi}_2(\infty) \,\rangle  &=\frac{|x|}{|x_\perp|^3}\,\left[(1+2\,\delta^{\alpha1})\,\hat{h}^2\,c(\widehat{\Phi}_3^{\alpha},\widehat{\Phi}_1^{\,\beta},\widehat{\Phi}_2)\,G_1^{-1}(\upsilon)\right.\\
       &\qquad\qquad\qquad\left.+ c(\Phi_3^{\alpha},\widehat{\Phi}_1^{\,\beta},\widehat{\Phi}_2)\, \sum_{n=0}^{\infty}\,\frac{(-1)^n\,(2)_n}{(n+5/2)_n}\,G_{2n+3}^{-1}(\upsilon)\right]  \ .
    \end{aligned}
\end{align}
Comparing \eqref{eq:conformal block Phi3} with \eqref{eq:conformal block expansion main}, we find that the coefficients associated with $\widehat{\mathsf{O}}_{2n+3}^{\,\alpha}$ must satisfy the equalities \eqref{eq:btd 2n+3 OPE coeff}.

\section{DCFT data on the line defect}\label{eq:DCFT data on the line defect}
We are now in a position to apply to the critical line defect the axiomatic method described in section \ref{eq:DCFT axiom} and read off the conformal dimensions of several defect composite operators as the DCFT data.

\subsection{Lowest-lying defect local operator and critical defect coupling}\label{1st order: scalar}
We consider the lowest-lying defect local operator $\widehat{W}_1^{\,\alpha}$, which reduces to $\widehat{\Phi}_1^{\,\alpha}$ in the $\epsilon\to0$ limit.
We assume that the critical defect coupling $\hat{h}$ is an $O(\epsilon^0)$ parameter to be fixed by defect conformal symmetry.
At the zero-th order in $\epsilon$, the conformal dimension of $\widehat{W}_1^{\,\alpha}$ should be
\begin{align}
   \widehat{\Delta}_{\widehat{W}_1^\alpha}=1+O(\epsilon)\ .
\end{align}
We expect that the symmetry breaking $\mathrm{O}(N)\rightarrow\mathrm{O}(N-1)$ on the defect makes $ \widehat{\Delta}_{\widehat{W}^{1}}$ and $ \widehat{\Delta}_{\widehat{W}^{\hat{\alpha}}}$ different at the first order in $\epsilon$.

We first apply Axiom \ref{dcftaxiom1} to fix the DOE of $W^\alpha_1$ as
\begin{align}\label{eq:V1 OPE Wilsonline}
    W^\alpha_1(x)\supset C_0^\alpha\,\frac{1}{|x_\perp|^{ \Delta_{W_1}}} \,\bm{1}+C_{1}^\alpha\,\frac{1}{|x_\perp|^{ \Delta_{W_1}- \widehat{\Delta}_{\widehat{W}_1^\alpha}}}\,\widehat{W}_1^{\,\alpha}(\hat{x}) \ .
\end{align}
For \eqref{eq:V1 OPE Wilsonline} to be identified with \eqref{eq:phi1 OPE Wilsonline} in the $\epsilon\to 0$ limit (Axiom \ref{dcftaxiom2}), we have
\begin{align}
    C_{1}^\alpha=1+O(\epsilon)\ ,\qquad
     C_{0}^\alpha=\delta^{\alpha1}\,\hat{h}+O(\epsilon) \ .
\end{align}
Combining all these, we now derive the leading anomalous dimension of $\widehat{W}_1^{\,\alpha}$.
The equation of motion \eqref{eq:classical EoM O(N)} with the DOE \eqref{eq:V1 OPE Wilsonline} (and the identity \eqref{eq:action of Lap on spinning btd OPE}) yields
\begin{align}
\begin{aligned}
        W^\alpha_3(x)
        &\supset\frac{C_{0}^\alpha}{\kappa}\,\frac{ \Delta_{W_1}\,( \Delta_{W_1}-1+\epsilon)}{|x_\perp|^{ \Delta_{W_1}+2}} \,\bm{1}  \\
       &\qquad\qquad+\frac{C_{1}^\alpha}{\kappa}\,\frac{( \Delta_{W_1}- \widehat{\Delta}_{\widehat{W}_1^\alpha})( \Delta_{W_1}- \widehat{\Delta}_{\widehat{W}_1^\alpha}-1+\epsilon)}{|x_\perp|^{ \Delta_{W_1}- \widehat{\Delta}_{\widehat{W}_1^\alpha}+2}}\,\widehat{W}_1^{\,\alpha}(\hat{x})\ .\\
\end{aligned}
\end{align}
Axiom \ref{dcftaxiom2} requires that this DOE should match \eqref{eq:phi3 OPE Wilsonline} in the $\epsilon\rightarrow 0$ limit:
\begin{align}
\hat{h}^3\,\kappa&=\hat{h}\, \Delta_{W_1}\,( \Delta_{W_1}-1+\epsilon)+O(\epsilon^2) \ ,\label{eq:equation for defect coupling}\\
(1+2\,\delta^{\alpha1})\,\hat{h}^2\,\kappa&=( \Delta_{W_1}- \widehat{\Delta}_{\widehat{W}_1^{\alpha}})( \Delta_{W_1}- \widehat{\Delta}_{\widehat{W}_1^{\alpha}}-1+\epsilon)+O(\epsilon^2) \ .\label{eq:equation for hat W1}
\end{align}
With the bulk parameters \eqref{eq:input from bulk} substituted, the first equation \eqref{eq:equation for defect coupling} gives the critical value of the defect coupling:
\begin{align}\label{eq:aphi in O(N) model WFFP}
    \hat{h}^2=\frac{N+8}{4}+O(\epsilon)\ .
\end{align}
Plugging this result into the second equation \eqref{eq:equation for hat W1}, we obtain the conformal dimension:
\begin{align}\label{conf dim: defect local scalar}
    \begin{aligned}
        \widehat{\Delta}_{\widehat{W}_1^{\alpha}}
            &=
                \Delta_{W_1}+\frac{1+2\,\delta^{\alpha1}}{2}\,\epsilon+O(\epsilon^2) \\
            &=
                1+\epsilon\,\delta^{\alpha1}+O(\epsilon^2)\ ,
    \end{aligned}
\end{align}
which agrees with the diagrammatic results at order $O(\epsilon)$ \cite[equation (3.19) and (3.21)]{Cuomo:2021kfm}.

\subsection{Defect local operators with transverse spin}\label{1st order: transverse spin}
Next, we consider defect local operators with transverse spin, $\widehat{U}_{i_1\cdots i_s}^{\,1}$ and $\widehat{U}_{i_1\cdots i_s}^{\,\hat{\alpha}}$, of conformal dimension $\widehat{\Delta}_{\widehat{U}^{1}_s}$ and $\widehat{\Delta}_{\widehat{U}^{\hat{\alpha}}_s}$, respectively.
In the free theory limit, they reduce to the free theory operators as follows:
\begin{align}
   \lim_{\epsilon\to0}\,\widehat{U}_{i_1\cdots i_s}^{\,1}=\widehat{\Phi}_{s+1,i_1\cdots i_s}^{\,1}\ ,\qquad
   \lim_{\epsilon\to0}\,\widehat{U}_{i_1\cdots i_s}^{\,\hat{\alpha}}=   \widehat{\Phi}_{s+1,i_1\cdots i_s}^{\,\hat{\alpha}} \ .
\end{align}

The defect conformal symmetry restricts the form of the DOE of $W_1^\alpha$ to the defect operators with spin as
\begin{align}
        W^\alpha_1(x)\supset C_s^\alpha\,\frac{x_\perp^{(i_1}\cdots x_\perp^{i_s)}}{|x_\perp|^{ \Delta_{W_1}- \widehat{\Delta}_{\widehat{U}^{\alpha}_s}+s}}\,\widehat{U}_{i_1\cdots i_s}^{\,\alpha}(\hat{x})\ .
\end{align}
Compared with the DOE \eqref{eq:phi1 OPE Wilsonline} in the free limit, the coefficient $C_s^\alpha$ must be 
\begin{align}
C_s^\alpha=\frac{1}{s!}+O(\epsilon)\ .   
\end{align}
Using the equation of motion \eqref{eq:classical EoM O(N)} together with \eqref{eq:action of Lap on spinning btd OPE}, one has
\begin{align}
\begin{aligned}    W^\alpha_3(x)\supset\frac{C_s^\alpha}{\kappa}\,( \Delta_{W_1}- \widehat{\Delta}_{\widehat{U}^{\alpha}}+s)( \Delta_{W_1}- \widehat{\Delta}_{\widehat{U}^{\alpha}}-s-1+\epsilon)\,
              \frac{x_\perp^{(i_1}\cdots x_\perp^{i_s)}}{|x_\perp|^{ \Delta_{W_1}- \widehat{\Delta}_{\widehat{U}^{\alpha}_s}+s+2}}\,\widehat{U}_{i_1\cdots i_s}^{\,\alpha}(\hat{x})\ .    
\end{aligned}
\end{align}
This DOE should reduce to \eqref{eq:phi3 OPE Wilsonline} in the free limit.
Repeating a similar analysis to the last section, we arrive at
\begin{align}\label{eq:conf dim:transverse spin}
    \begin{aligned}
        \widehat{\Delta}_{\widehat{U}^{\alpha}_s}
            &= 
            \Delta_{W_1}+s+\frac{1+2\,\delta^{\alpha1}}{2\,(s+2)}\,\epsilon+O(\epsilon^2) \\
            &=
            s+1+\frac{2\,\delta^{\alpha1}-s-1}{2\,(s+2)}\,\epsilon+O(\epsilon^2)\ .
    \end{aligned}
\end{align}
We notice that the above result \eqref{eq:conf dim:transverse spin} is in agreement with the universal behavior of defect local operators with large transverse spins \cite{Lemos:2017vnx}:
\begin{align}\label{eq:large spin}        \widehat{\Delta}_{\widehat{U}^{\alpha}_s} \simeq
                \Delta_{W_1} + s\, ,  \qquad s\rightarrow \infty\ .
\end{align}

\subsection{Defect composite operators}\label{sec:defect composite operator}
Let us move on to the composite operators which tend to $\widehat{\Phi}_2$ listed in \eqref{eq:second order defect local app} in the free limit:
\begin{align}\label{eq:second order operators with no transverse spin indices}
 \widehat{W}_2 \in \left\{\widehat{V}^{\hat{\alpha}}\ ,~ \widehat{T}^{\hat{\alpha}\hat{\beta}}\ ,~ \widehat{S}_+\ , ~ \widehat{S}_-\right\} \ .
\end{align}
We denote the conformal dimension of $\widehat{W}_2$ by $\widehat{\Delta}_{\widehat{W}_2}$ and focus on its leading correction $\Gamma_{\widehat{W}_2}$:
\begin{align}
    \widehat{\Delta}_{\widehat{W}_2}
        =
            2+\Gamma_{\widehat{W}_2}\,\epsilon+O(\epsilon^2)\ .
\end{align}
The first two operators $\widehat{V}^{\hat{\alpha}}$ and $\widehat{T}^{\hat{\alpha}\hat{\beta}}$ are an O$(N-1)$ vector and a symmetric traceless tensor, respectively. 
In the free theory limit,
\begin{align}
    \lim_{\epsilon\to0}\,\widehat{V}^{\hat{\alpha}}
        =
        \widehat{\Phi}_{1}^{\,1}\widehat{\Phi}_{1}^{\,\hat{\alpha}}\ ,\qquad
    \lim_{\epsilon\to0}\,\widehat{T}^{\hat{\alpha}\hat{\beta}}
        =
        \widehat{\Phi}_{1}^{\,(\hat{\alpha}}\widehat{\Phi}_{1}^{\,\hat{\beta})}\ .
\end{align}
The last two operators $\widehat{S}_+$ and $\widehat{S}_-$ are O$(N-1)$ scalars and can be identified as linear combinations of $|\widehat{\Phi}_1^{\,1}|^2$ and $|\widehat{\Phi}_1^{\,\hat{\gamma}}|^2$ in the free limit.\footnote{There is no operator mixing between $\hat S_\pm$ and the other operators ($\widehat{V}^{\hat{\alpha}}$ and $\widehat{T}^{\hat{\alpha}\hat{\beta}}$) as they are in different irreducible representations of the symmetry group \eqref{eq:residual symmetry group on the defect}.}
It is convenient to use the following parametrization:
\begin{align}\label{eq:scalar mixing}
    \lim_{\epsilon\to0}\begin{pmatrix}
    \widehat{S}_+\\ \widehat{S}_-
    \end{pmatrix}
    =
    \begin{pmatrix}
    \cos\theta & -\sin\theta\\
    \sin\theta &\cos\theta
    \end{pmatrix}\,
    \begin{pmatrix}
    \frac{1}{\sqrt{2}}\,|\widehat{\Phi}_1^{\,1}|^2\\ \frac{1}{\sqrt{2\,(N-1)}}\,|\widehat{\Phi}_1^{\,\hat{\gamma}}|^2
    \end{pmatrix}\ ,
\end{align}
so that two-point functions of $\widehat{S}_\pm$ are unit-normalized and are orthogonal to each other.
We take the same strategy as \cite[section 5.2]{BCFTpaper} to calculate the conformal dimensions of these composite operators.

Firstly, we employ the equation of motion \eqref{eq:classical EoM O(N)} to determine the DOE of $W_1^{\alpha}$ at the first order in $\epsilon$.
It turns out that a series of operators $\widehat{\mathsf{O}}_{2n+3}^{\prime\,\alpha}$ with conformal $\widehat{\Delta}'_{2n+3}=2n+3+O(\epsilon)$ that reduce to $\widehat{\mathsf{O}}_{2n+3}$ in the free limit appear in the DOE of $W^\alpha_1$:\footnote{Due to the symmetry breaking on the defect O$(N)\to$ O$(N-1)$, the conformal multiplet of $\widehat{\mathsf{O}}_{2n+3}^{\,\alpha}$ in general
splits into two parts $\widehat{\mathsf{O}}_{2n+3}^{\prime\,1}$ and $\widehat{\mathsf{O}}_{2n+3}^{\prime\,\hat{\alpha}}$, and their conformal dimensions are different at order $O(\epsilon)$.
Nevertheless, such differences in conformal dimensions do not affect our arguments and can be ignored in subsequent discussions.}
\begin{align}\label{eq:OPE of W1 all order}
    \begin{aligned}
        W_1^\alpha(x)
            &\supset
                \frac{C_1^\alpha}{|x_\perp|^{\Delta_1-\widehat{\Delta}_{\widehat{W}_{1}^{\, \alpha}}}}\,\widehat{W}_1^{\,\alpha}(\hat{x})
                +
                \sum_{n=0}^{\infty}\,\frac{b(W_1^\alpha,\widehat{\mathsf{O}}_{2n+3}^{\prime\,\alpha})/c(\widehat{\mathsf{O}}_{2n+3}^{\prime\,\alpha},\widehat{\mathsf{O}}_{2n+3}^{\prime\,\alpha})}{|x_\perp|^{\Delta_1-\widehat{\Delta}'_{2n+3}}}\,\widehat{\mathsf{O}}_{2n+3}^{\prime\,\alpha}(\hat{x})\ .
    \end{aligned}
\end{align}
$C_1^\alpha=1+O(\epsilon)$ is the same constant as in section \ref{1st order: scalar}.
By applying the equation of motion \eqref{eq:classical EoM O(N)}, we have
{\small\begin{align}
        W_3^\alpha(x)
            \supset
                \sum_{n=0}^{\infty}\,\frac{b(W_1^\alpha,\widehat{\mathsf{O}}_{2n+3}^{\prime\,\alpha})}{\kappa\,c(\widehat{\mathsf{O}}_{2n+3}^{\prime\,\alpha},\widehat{\mathsf{O}}_{2n+3}^{\prime\,\alpha})}\,\frac{(\Delta_1-\widehat{\Delta}'_{2n+3})(\Delta_1-\widehat{\Delta}'_{2n+3}-1+\epsilon)}{|x_\perp|^{\Delta_1-\widehat{\Delta}'_{2n+3}+2}}\,\widehat{\mathsf{O}}_{2n+3}^{\prime\,\alpha}(\hat{x}) \ .
\end{align}}
Taking $\epsilon\to0$ limit and comparing with \eqref{eq:phi3 OPE Wilsonline}, we find
\begin{align}\label{eq:W1 to higher order OPE coeff}
    \frac{b(W_1^\alpha,\widehat{\mathsf{O}}_{2n+3}^{\prime\,\alpha})}{c(\widehat{\mathsf{O}}_{2n+3}^{\prime\,\alpha},\widehat{\mathsf{O}}_{2n+3}^{\prime\,\alpha})}=\frac{\kappa}{2\,(n+1)(2n+3)}\,\frac{b(\Phi_3^\alpha,\widehat{\mathsf{O}}_{2n+3}^{\,\alpha})}{c(\widehat{\mathsf{O}}_{2n+3}^{\,\alpha},\widehat{\mathsf{O}}_{2n+3}^{\,\alpha})}+O(\epsilon^2)\ .
\end{align}
With the DOE \eqref{eq:OPE of W1 all order} of $W_1^\alpha$, the three-point function $\langle\, W_1^{\,\alpha}\,\widehat{W}_1^{\,\beta}\,\widehat{W}_2 \,\rangle$ can be calculated as
\begin{align}\label{eq:3pt app W1}
    \begin{aligned}
    \langle\,& W_1^\alpha(x)\, \widehat{W}_1^{\,\beta}(0)\,\widehat{W}_{2}(\infty)\,\rangle
        =
            \frac{1}{|x_\perp|^{\Delta_1}\,|x|^{\widehat{\Delta}_{\widehat{W}_{1}^{\, \beta}}-\widehat{\Delta}_{\widehat{W}_2}}}\\
        &\qquad\cdot
            \left[C_1^\alpha\cdot c(\widehat{W}_1^{\,\alpha},\widehat{W}_1^{\,\beta},\widehat{W}_2)\,G^{\widehat{\Delta}_{\widehat{W}_{1}^{\, \beta}}-\widehat{\Delta}_{\widehat{W}_2}}_{\widehat{\Delta}_{\widehat{W}_{1}^{\, \alpha}}}\left(\frac{|x_\perp|^2}{|x|^2}\right)\right.\\
        &\qquad\qquad\left.     
            +
            \sum_{n=0}^{\infty}\,\frac{b(W_1^\alpha,\widehat{\mathsf{O}}_{2n+3}^{\prime\,\alpha})\,c(\widehat{\mathsf{O}}_{2n+3}^{\prime\,\alpha},\widehat{W}_1^{\,\beta},\widehat{W}_2)}{c(\widehat{\mathsf{O}}_{2n+3}^{\prime\,\alpha},\widehat{\mathsf{O}}_{2n+3}^{\prime\,\alpha})}\,G_{\widehat{\Delta}'_{2n+3}}^{\widehat{\Delta}_{\widehat{W}_{1}^{\, \beta}}-\widehat{\Delta}_{\widehat{W}_2}}\left(\frac{|x_\perp|^2}{|x|^2}\right)\right]\ .
    \end{aligned}
\end{align}
With \eqref{conf dim: defect local scalar}, we expand the first term in the parenthesis in powers of $\epsilon$ as:
\begin{align}\label{eq: G}
\begin{aligned}
    G^{\widehat{\Delta}_{\widehat{W}_{1}^{\, \beta}}-\widehat{\Delta}_{\widehat{W}_2}}_{\widehat{\Delta}_{\widehat{W}_{1}^{\, \alpha}}}(\upsilon)&=\upsilon^{\widehat{\Delta}_{\widehat{W}_{1}^{\, \alpha}}/2}\,{}_2F_1\left(\tfrac{\delta^{\alpha1}+\delta^{\beta1}-\Gamma_{\widehat{W}_2}}{2}\,\epsilon,1;3/2;\upsilon\right)+O(\epsilon^2)\\
    &=\upsilon^{\widehat{\Delta}_{\widehat{W}_{1}^{\, \alpha}}}+\frac{\delta^{\alpha1}+\delta^{\beta1}-\Gamma_{\widehat{W}_2}}{3}\,\epsilon\,\upsilon^{3/2}\,{}_2F_1(1,1;5/2;\upsilon)+O(\epsilon^2)\ .
\end{aligned}
\end{align}
On the other hand, the second term turns out to be\footnote{We used \eqref{eq:W1 to higher order OPE coeff} and \eqref{eq:btd 2n+3 OPE coeff}, and the sum rule for the hypergeometric function:
\begin{align*}
    {}_2F_1(1,1;5/2;z)=\sum_{n=0}^{\infty}\,\frac{(-1)^n\,n!}{(2n/3+1)\,(n+5/2)_n} \, z^{n}\,{}_2F_1(n+1,n+2;2n+7/2;z)\ ,
\end{align*}
which can be proved by expanding ${}_2F_1$ in the RHS and rearranging in powers of $z$.
}
\begin{align}\label{eq: sum G}
    \begin{aligned}
        \sum_{n=0}^{\infty}\,\frac{b(W_1^\alpha,\widehat{\mathsf{O}}_{2n+3}^{\prime\,\alpha})\,c(\widehat{\mathsf{O}}_{2n+3}^{\prime\,\alpha},\widehat{W}_1^{\,\beta},\widehat{W}_2)}{c(\widehat{\mathsf{O}}_{2n+3}^{\prime\,\alpha},\widehat{\mathsf{O}}_{2n+3}^{\prime\,\alpha})}\,G_{\widehat{\Delta}'_{2n+3}}^{\widehat{\Delta}_{\widehat{W}_{1}^{\, \beta}}-\widehat{\Delta}_{\widehat{W}_2}}&(\upsilon)\\
        = \frac{\kappa\, c(\Phi_3^{\alpha},\widehat{\Phi}_1^{\,\beta},\widehat{\Phi}_2)}{6}\,&\upsilon^{3/2}\,{}_2F_1(1,1;5/2;\upsilon)+O(\epsilon^2)\ .
    \end{aligned}
\end{align}
By plugging \eqref{eq: G} and \eqref{eq: sum G} into \eqref{eq:3pt app W1}, we end up with\footnote{We use the following relation:
\begin{align*}
     C_1^\alpha\cdot c(\widehat{W}_1^{\,\alpha},\widehat{W}_1^{\,\beta},\widehat{W}_2)=c(\widehat{\Phi}_1^{\,\alpha},\widehat{\Phi}_1^{\,\beta},\widehat{\Phi}_2)+O(\epsilon)\ .
\end{align*}}
\begin{align}\label{eq:3pt gen}
\begin{aligned}
     \langle\, &W^{\alpha}_1(x)\,\widehat{W}_1^{\,\beta}(0)\,\widehat{W}_2(\infty) \,\rangle\\
     &= c(\widehat{W}_1^{\,\alpha},\widehat{W}_1^{\,\beta},\widehat{W}_2)\,\frac{C_1^\alpha}{|x_\perp|^{\Delta_{W_1}-\widehat{\Delta}_{\widehat{W}_1^{\beta}}} \,|x|^{\widehat{\Delta}_{\widehat{W}_{1}^{\alpha}}+\widehat{\Delta}_{\widehat{W}_1^{\beta}}-\widehat{\Delta}_{\widehat{W}_2}}}\\
     &\qquad\qquad + \frac{\epsilon}{3\,(N+8)}\,\left[(N+8)(\delta^{\alpha1}+\delta^{\beta1}-\Gamma_{\widehat{W}_2})\,c(\widehat{\Phi}_1^{\,\alpha},\widehat{\Phi}_1^{\,\beta},\widehat{\Phi}_2)+c(\widehat{\Phi}_3^{\,\alpha},\widehat{\Phi}_1^{\,\beta},\widehat{\Phi}_2)\right]\\
     &\qquad\qquad\qquad\qquad\qquad\qquad\qquad\qquad\cdot \frac{|x_\perp|^2}{|x|^2}\,{}_2F_1\left(1,1;\frac{5}{2};\frac{|x_\perp|^2}{|x|^2}\right)+O(\epsilon^2)\ .
\end{aligned}
\end{align}
Let us take $\hat{x}\to0$ in \eqref{eq:3pt gen} where the bulk operator is still distant from defects.
We observe \eqref{eq:3pt gen} is not analytic in the limit due to the odd integer powers of $|\hat{x}|$.
This non-analytic behavior originates from the asymptotic form of the Gauss's hypergeometric function:\footnote{Use Kummer's connection formula for hypergeometric functions:
\begin{align*}
    \begin{aligned}
    {}_2F_1(\alpha,&\beta;\gamma;z)=\frac{\Gamma(\gamma)\Gamma(\gamma-\alpha-\beta)}{\Gamma(\gamma-\alpha)\Gamma(\gamma-\beta)}\,{}_2F_1(\alpha,\beta;\alpha+\beta-\gamma+1;1-z)\\
    &\quad +\frac{\Gamma(\gamma)\Gamma(\alpha+\beta-\gamma)}{\Gamma(\alpha)\Gamma(\beta)}\,(1-z)^{\gamma-\alpha-\beta}\,{}_2F_1(\gamma-\alpha,\gamma-\beta;\gamma-\alpha-\beta+1;1-z)\ .
    \end{aligned}
\end{align*}
}
\begin{align}
    {}_2F_1\left(1,1;\frac{5}{2};\frac{|x_\perp|^2}{|x|^2}\right)\xrightarrow[\hat{x}\sim0]{}\frac{3\,\pi}{2}\cdot \frac{|\hat{x}|}{|x|}+\cdots\ .
\end{align}
which contradicts the holomorphy of Euclidean correlators away from the coincidence of points.
The only way to resolve this tension is to set the coefficient in front of the singular term in \eqref{eq:3pt gen} to zero:
\begin{align}\label{eq:master equation}
    (N+8)\,(\delta^{\alpha1}+\delta^{\beta1}-\Gamma_{\widehat{W}_2})\,c(\widehat{\Phi}_1^{\,\alpha},\widehat{\Phi}_1^{\,\beta},\widehat{\Phi}_2)+c(\widehat{\Phi}_3^{\,\alpha},\widehat{\Phi}_1^{\,\beta},\widehat{\Phi}_2)=0 \ .
\end{align}
By plugging the defect three-point coefficients $c(\widehat{\Phi}_1^{\alpha},\widehat{\Phi}_1^{\,\beta},\widehat{\Phi}_2)$ and $c(\widehat{\Phi}_3^{\alpha},\widehat{\Phi}_1^{\,\beta},\widehat{\Phi}_2)$ listed in table \ref{tab:list three-point coeff} into the above equation, we obtain the conformal dimension of the operators at the first order in $\epsilon$ together with the mixing angle $\theta$ of the scalar operators \eqref{eq:scalar mixing}:\footnote{The conformal dimensions of $\widehat{\Delta}_{\widehat{S}_\pm}$ and the mixing angle $\theta$ are derived by solving the following simultaneous equations:
\begin{align*}
\begin{aligned}
 (N+8)\, (\Gamma_{\widehat{S}_+}&-\delta^{\alpha1}-\delta^{\beta1})\, \left[\sqrt{N-1}\,c(\widehat{\Phi}_1^{\,\alpha},\widehat{\Phi}_1^{\,\beta},|\widehat{\Phi}_1^{\,1}|^2)-\tan\theta\,c(\widehat{\Phi}_1^{\,\alpha},\widehat{\Phi}_1^{\,\beta},|\widehat{\Phi}_1^{\,\hat{\gamma}}|^2)\right]\\
  &=\sqrt{N-1}\,c(\widehat{\Phi}_3^{\,\alpha},\widehat{\Phi}_1^{\,\beta},|\widehat{\Phi}^{\,1}|^2)-\tan\theta\,c(\widehat{\Phi}_3^{\,\alpha},\widehat{\Phi}_1^{\,\beta},|\widehat{\Phi}_1^{\,\hat{\gamma}}|^2)   \ ,\\
 (N+8)\, (\Gamma_{\widehat{S}_-}&-\delta^{\alpha1}-\delta^{\beta1})\, \left[\sqrt{N-1}\,\tan\theta\,c(\widehat{\Phi}^{\,\alpha},\widehat{\Phi}_1^{\,\beta},|\widehat{\Phi}^{\,1}|^2)+c(\widehat{\Phi}^{\,\alpha},\widehat{\Phi}_1^{\,\beta},|\widehat{\Phi}^{\,\hat{\gamma}}|^2)\right]\\
  &=\sqrt{N-1}\,\tan\theta\,c(\widehat{\Phi}_3^{\,\alpha},\widehat{\Phi}_1^{\,\beta},|\widehat{\Phi}^{\,1}|^2)+c(\widehat{\Phi}_3^{\,\alpha},\widehat{\Phi}_1^{\,\beta},|\widehat{\Phi}^{\,\hat{\gamma}}|^2)    \ .
\end{aligned}
\end{align*}
There are two sets of solutions and we have chosen the ones satisfying $\Gamma_{\widehat{S}_+}\geq \Gamma_{\widehat{S}_-}$.
}
\begin{align}
\widehat{\Delta}_{\widehat{V}}&=2+\frac{N+10}{N+8}\,\epsilon+O(\epsilon^2)\ ,\label{conf dim: vector}\\
\widehat{\Delta}_{\widehat{T}}&=2+\frac{2}{N+8}\,\epsilon+O(\epsilon^2)\ , \label{conf dim: tensor}\\
\widehat{\Delta}_{\widehat{S}_\pm}&=2+\frac{3N+20\pm \sqrt{N^2+40N+320}}{2\,(N+8)}\,\epsilon+O(\epsilon^2)\ ,\label{conf dim: scalars}\\
\tan\theta&=\frac{N+18+\sqrt{N^2+40N+320}}{2\,\sqrt{N-1}}\ .
\end{align}

\subsection{Defect operator spectrum for $N=1$ (Ising DCFT)}\label{sec:N=1 anomalous dimension}
Finally, consider the case with $N=1$ (Ising CFT with a localized magnetic field).
Note that there are no flavor symmetries, hence no symmetry breaking on the defect. 
We can compute the anomalous dimensions of the defect operators $\widehat{W}_p(\hat{x})$ that tend to $\widehat{\Phi}_{p}(\hat{x})$ in $\epsilon\to 0$ limit. (Recall \eqref{eq: def of p-order compodite op}.)

Performing a similar analysis to the last subsection for $\langle\,W_1\,\widehat{W}_{p}\,\widehat{W}_{p+1}\,\rangle$ leads
\begin{align}\label{eq:master equation N=1}
    9\,(\Gamma_{\widehat{W}_1}+\Gamma_{\widehat{W}_{p}}-\Gamma_{\widehat{W}_{p+1}})\,c(\widehat{\Phi}_1,\widehat{\Phi}_p,\widehat{\Phi}_{p+1})+c(\widehat{\Phi}_3,\widehat{\Phi}_p,\widehat{\Phi}_{p+1})=0\ ,
\end{align}
with 
\begin{align}
    c(\widehat{\Phi}_1,\widehat{\Phi}_p,\widehat{\Phi}_{p+1})=(p+1)!\ ,\qquad c(\widehat{\Phi}_3,\widehat{\Phi}_p,\widehat{\Phi}_{p+1})=3p\,(p+1)!\ ,
\end{align}
and
\begin{align}
    \widehat{\Delta}_{\widehat{W}_{p}}=p+\Gamma_{\widehat{W}_p}\,\epsilon+O(\epsilon^2)\ .
\end{align}
\eqref{conf dim: defect local scalar} gives $\Gamma_{\widehat{W}_1}=1$.
By solving the recursion relation $\Gamma_{\widehat{W}_{p+1}}=\Gamma_{\widehat{W}_p}+1+p/3$ that follows from \eqref{eq:master equation N=1}, we obtain
\begin{align}
   \widehat{\Delta}_{\widehat{W}_{p}}=p+\frac{p\,(p+5)}{6}\,\epsilon+O(\epsilon^2)\ .
\end{align}
Note that this is consistent with \eqref{conf dim: defect local scalar} and the $O(\epsilon)$ correction to the conformal dimension of $\widehat{W}_2$ is identical to that of $\widehat{S}_+$ with $N=1$.

\acknowledgments
The work of T.\,N. was supported in part by the JSPS Grant-in-Aid for Scientific Research (C) No.19K03863, Grant-in-Aid for Scientific Research (A) No.\,21H04469, and
Grant-in-Aid for Transformative Research Areas (A) ``Extreme Universe''
No.\,21H05182 and No.\,21H05190.
The work of Y.\,O. was supported by Forefront Physics and Mathematics Program to Drive Transformation (FoPM), a World-leading Innovative Graduate Study (WINGS) Program, the University of Tokyo.
The work of Y.\,O. was also supported by JSPS fellowship for young students, MEXT, and by JSR fellowship, the University of Tokyo.

\appendix

\section{Bulk-defect-defect three-point function}\label{app:bdd 3pt}
Following \cite[appendix C]{Lauria:2020emq}, we expand on the scalar bulk-defect-defect three-point functions in DCFT.\footnote{The readers interested in more on group theoretical perspectives are also referred to \cite{Buric:2020zea}. A similar analysis to \cite{Lauria:2020emq} has been carried out in \cite{Karch:2018uft} for boundary CFT, which fails to choose the proper solution to the conformal Casimir equation.} In the same manner as section \ref{eq:Structures of DCFT}, we let the spacetime dimensions $d$ and defect dimensions $p$ be general for future reference.

We first introduce the embedding space formalism for DCFT (appendix \ref{app:Notations}) and derive the conformal block expansion for scalar bulk-defect-defect three-point function by solving conformal Casimir equation (appendix \ref{app:Conformal block expansion}). Then, we apply the DOEs inside the three-point function in appendix \ref{app:Reconstruction of conformal block}. After summing over DOEs for all orders, we confirm that it reproduces the conformal block expansion derived in appendix \ref{app:Conformal block expansion}.
\subsection{Embedding space formalism}\label{app:Notations}
We embed the physical coordinate system $\mathbb{R}^d$ onto the projective null cone on $X^M\in\mathbb{R}^{1,d+1}$ with $M,N=-,+,1,\cdots, d$ \cite{Costa:2011mg}:
\begin{align}
\begin{aligned}
           \d s_{\mathbb{R}^{1,d+1}}^2&=\eta_{MN}\,\d X^M\d X^N=-\d X^+ \d X^- +\delta_{\mu\nu}\,\d X^\mu \d X^\nu\ ,\\
         \delta_{\mu\nu}=\text{diag}&(1,\cdots,1)\ ,\qquad  \eta_{MN}\,X^M X^N=0\ ,\qquad X^M\sim \lambda\,X^M\ ,\qquad \lambda\in\mathbb{R}_+\ .
\end{aligned}
\end{align}
We denote the embedding space coordinates on and around the defect by $Q^M=(Q^A,Q^I=0)$ and $P^M=(P^A,P^I)$ with $A=+,-,1,\cdots,p$ and $I=p+1,\cdots,d$. In going back to the physical coordinates, one should make the following replacements:
\begin{align}
    P^M=(P^+,P^-,P^a,P^i)\mapsto(1,x^2,\hat{x}^a,x_\perp^i)\ ,\qquad  Q^A=(Q^+,Q^-,Q^a)\mapsto(1,\hat{y}^2,\hat{y}^a)\ .
\end{align}
The bulk and defect local scalars in the embedding space $\CO_{\Delta}(P)$ and $\widehat{\CO}_{\widehat{\Delta}}(Q)$ are specified by the homogeneity associated with their conformal dimensions $\Delta$ and $\widehat{\Delta}$:
\begin{align}\label{eq:homogeneity}
P^M\,\frac{\partial}{\partial P^M}\,\CO_{\Delta}(P)=-\Delta\,\CO_{\Delta}(P)\ ,\qquad Q^A\,\frac{\partial}{\partial Q^A}\,\widehat{\CO}_{\widehat{\Delta}}(Q)=-\widehat{\Delta}\,\widehat{\CO}_{\widehat{\Delta}}(Q)\ .
\end{align}
The generators of defect conformal group $\mathbf{J}_{AB},\mathbf{J}_{IJ}$ act as rotation differential operators on the embedding space operators:
\begin{align}\label{eq:action of conf gen}
\begin{aligned}
          [ \mathbf{J}_{AB},\,\CO_{\Delta}(P)]&=-\CJ_{AB}(P)\,\CO_{\Delta}(P)=-\left(P_A\,\frac{\partial}{\partial P^B}-P_B\,\frac{\partial}{\partial P^A}\right)\,\CO_{\Delta}(P)\ ,\\
 [ \mathbf{J}_{IJ},\,\CO_{\Delta}(P)]&=-\CJ_{IJ}(P)\,\CO_{\Delta}(P)=-\left(P_I\,\frac{\partial}{\partial P^J}-P_J\,\frac{\partial}{\partial P^I}\right)\,\CO_{\Delta}(P)\ ,\\
           [ \mathbf{J}_{AB},\,\widehat{\CO}_{\widehat{\Delta}}(Q)]&=-\CJ_{AB}(Q)\,\widehat{\CO}_{\widehat{\Delta}}(Q)=-\left(Q_A\,\frac{\partial}{\partial Q^B}-Q_B\,\frac{\partial}{\partial Q^A}\right)\,\widehat{\CO}_{\widehat{\Delta}}(Q)\ ,\\
                      [ \mathbf{J}_{IJ},\,\widehat{\CO}_{\widehat{\Delta}}(Q)]&=-\CJ_{IJ}(Q)\,\widehat{\CO}_{\widehat{\Delta}}(Q)=-\left(Q_I\,\frac{\partial}{\partial Q^J}-Q_J\,\frac{\partial}{\partial Q^I}\right)\,\widehat{\CO}_{\widehat{\Delta}}(Q)\ ,\\
\end{aligned}
\end{align}
We also use the following shorthanded notations to express $\mathrm{SO}(1,p+1)\times\mathrm{SO}(d-p)$ invariant inner products:
\begin{align}
X\cdot X'=X^M X'_M\ ,\qquad X\bullet X'=X^AX'_A\ ,\qquad X\circ X'=X^I X'_I\ .
\end{align}
One should pay attention to the following relations:
\begin{align}
  P\cdot P=  P\bullet P+P\circ P=0\ ,\qquad Q\cdot Q=Q\bullet Q=0\ .
\end{align}

\subsection{Conformal block expansion}\label{app:Conformal block expansion}
\begin{figure}[ht!]
    \centering
       \begin{tikzpicture}
          \draw[very thick, orange, opacity=0.9] (0,-3)  -- (0,2.7);    
                  \node[below, orange] at (0,-3) {\large $\CD^{(p)}$};
        
        \coordinate[label=0:$\hat{y}_1$] (A) at (0,-2) {};
        \coordinate[label=180:$x$] (B) at (-2.6,-0.3) {};
        \coordinate[label=0:$\hat{y}_2$] (C) at (0,1.5) {};
        
        \path[draw] (A) -- (B) -- (C) -- cycle;
        \draw[fill=gray!15,opacity=0.8]    (A) -- (B) -- (C) -- cycle;
        \node[RoyalBlue!100,font=\large] at (-1,-0.3) {$S$};
        
        \path pic["$\varphi$",draw,angle radius=7mm,angle eccentricity=1.3] {angle = A--B--C};
     \end{tikzpicture} 
    \caption{The triangle spanned by one point $x$ on the bulk and two points $\hat{y}_1,\hat{y}_2$ on the defect $\CD^{(p)}$.}
    \label{fig:cross-ratio}
\end{figure}
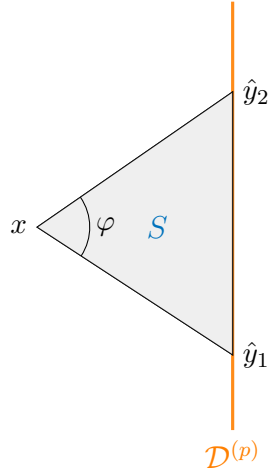

The three-point function of one bulk scalar $\CO_{\Delta_1}$ and two defect scalars $\widehat{\CO}_{\widehat{\Delta}_1},\widehat{\CO}_{\widehat{\Delta}_2}$ is fixed by defect conformal symmetry;
\begin{align}\label{eq:three-point embed}
\begin{aligned}
        \langle\, \CO_{\Delta}(P)&\,\widehat{\CO}_{\widehat{\Delta}_1}(Q_1)\,\widehat{\CO}_{\widehat{\Delta}_2}(Q_2) \,\rangle\\
        &=\frac{g(\upsilon)}{(P\circ P)^{\frac{\Delta}{2}} \,(-2P\bullet Q_1)^{\frac{\widehat{\Delta}^-_{12}}{2}}\,(-2P\bullet Q_2)^{\frac{\widehat{\Delta}^-_{21}}{2}}\,(-2Q_1\bullet Q_2)^{\frac{\widehat{\Delta}^+_{12}}{2}} }\ .
\end{aligned}
\end{align}
In the physical space, we have
\begin{align}\label{eq:three-point phys}
        \langle\, \CO_{\Delta}(x)\,\widehat{\CO}_{\widehat{\Delta}_1}(\hat{y}_1)\,\widehat{\CO}_{\widehat{\Delta}_2}(\hat{y}_2) \,\rangle=\frac{g(\upsilon)}{|x_\perp|^{\Delta} \,|x-\hat{y}_1|^{\widehat{\Delta}^-_{12}}\,|x-\hat{y}_2|^{\widehat{\Delta}^-_{21}}\,|\hat{y}_{12}|^{\widehat{\Delta}^+_{12}} }\ .
\end{align}
Here $\widehat{\Delta}^\pm_{ij}=\widehat{\Delta}_i\pm \widehat{\Delta}_j$ and $g(\upsilon)$ is some function of the $\mathrm{SO}(1,p+1)\times\mathrm{SO}(d-p)$ invariant $\upsilon$ defined by
\begin{align}\label{eq:three-point cross-ratio}
    \upsilon=\frac{(P\circ P)\,(-2Q_1\bullet Q_2)}{(-2P\bullet Q_1)\,(-2P\bullet Q_2)}\xrightarrow[\text{space}]{\text{physical}}\,\frac{|x_\perp|^2\,|\hat{y}_{12}|^2}{|x-\hat{y}_1|^2\,|x-\hat{y}_2|^2}
    \ .
\end{align}
Note that in the Euclidean regime, we have
\begin{align}\label{eq:upsilon range}
    0\leq \upsilon\leq1\ .
\end{align}
To see why this condition holds, consider the area of a triangle $S$ with the vertices at $x$, $\hat{y}_1$ and $\hat{y}_2$ (see figure \ref{fig:cross-ratio}). On one hand, we have $S=\frac{1}{2}\,|x_\perp|\,|\hat{y}_{12}|$, while using the angle $\varphi$ between $(x-y_1)$ and $(x-y_2)$, the area can be written by $S=\frac{1}{2}\,|\sin\varphi|\,|x-\hat{y}_1|\,|x-\hat{y}_2|$. Therefore, $ 2 S=|x_\perp|\,|\hat{y}_{12}|=|\sin\varphi|\,|x-\hat{y}_1|\,|x-\hat{y}_2|$ and 
$0\leq \upsilon=|\sin\varphi|^2\leq 1$.

\begin{figure}[ht!]
    \centering
       \begin{tikzpicture}
          \draw[very thick, orange, opacity=0.9] (0,-3)  -- (0,2.7);    
                  \node[below, orange] at (0,-3) {\large $\CD^{(p)}$};
        
        \coordinate[label=0:$\hat{y}_1$] (A) at (0,-2) {};
        \coordinate[label=180:$x$] (B) at (-1.6,-0.3) {};
        \coordinate[label=0:$\hat{y}_2$] (C) at (0,1.5) {};
                \filldraw[black,very thick] (A) circle (0.05);
                                \filldraw[black,very thick] (B) circle (0.05);
                                                \filldraw[black,very thick] (C) circle (0.05);
            \draw[black!100]  (0,-0.3) rectangle (-0.2,-0.1);           
                 \draw[black!100,dashed,opacity=0.8]  (-1.6,-0.3) -- (0,-0.3) ;           
        
\draw[dashed,very thick] (B) circle [radius=1.6];
     \end{tikzpicture} 
    \caption{The completeness relation is inserted on the dashed circle centered at $x$.}
    \label{fig:radial quantization}
\end{figure}
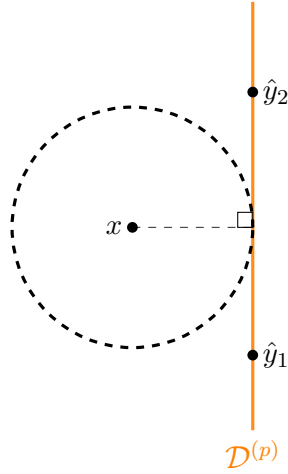

We now take the origin of the radial quantization at $x$, draw a circle tangent to the defect, and insert a completeness relation that diagonalizes the $\mathrm{SO}(1,p+1)\times\mathrm{SO}(d-p)$ Casimir (see figure \ref{fig:radial quantization}). 
Because the three-point function \eqref{eq:three-point embed} and cross-ratio \eqref{eq:three-point cross-ratio} are $\mathrm{SO}(d-p)$ singlet, we have only to consider the $\mathrm{SO}(1,p+1)$ part of the Casimir.
In what follows, we consider the $\mathrm{SO}(1,p+1)$ eigenstates that are spanned by the defect local scalars, namely
\begin{align}\label{eq:block exp 1}
             \langle\, \CO_{\Delta}(P)\,\widehat{\CO}_{\widehat{\Delta}_1}(Q_1)\,\widehat{\CO}_{\widehat{\Delta}_2}(Q_2) \,\rangle&=\sum_{\widehat{\Delta}}\, \langle\widehat{\CD}|\,\CR\{ \widehat{\CO}_{\widehat{\Delta}_1}(Q_1)\,\widehat{\CO}_{\widehat{\Delta}_2}(Q_2)\}\,|\,\widehat{\CO}_{\widehat{\Delta}}\,|\, \CO_{\Delta}(P)\,|\Omega\rangle\ ,
\end{align}
where $\CR$ means the appropriate radial ordering and $\langle\widehat{\CD}|$ is the defect vacuum.
Then, from \eqref{eq:homogeneity} and \eqref{eq:action of conf gen} the Casimir eigenvalue for $|\,\widehat{\CO}_{\widehat{\Delta}}\,|$ is given by\footnote{This comes from $\frac{1}{2}\,\mathbf{J}^{AB}\mathbf{J}_{AB}\,\widehat{\CO}_{\widehat{\Delta}}(Q)=-Q\bullet\,\frac{\partial}{\partial Q}\,\left(p+Q\bullet\,\frac{\partial}{\partial Q}\right)\,\widehat{\CO}_{\widehat{\Delta}}(Q)=-\widehat{\Delta}\,(\widehat{\Delta}-p)\,\widehat{\CO}_{\widehat{\Delta}}(Q)$.}
\begin{align}\label{eq:eigenvector of defect conformal group}
\frac{1}{2}\,\mathbf{J}^{AB}\mathbf{J}_{AB}\,|\,\widehat{\CO}_{\widehat{\Delta}}\,|=-\widehat{\Delta}\,(\widehat{\Delta}-p)\,|\,\widehat{\CO}_{\widehat{\Delta}}\,|\ .
\end{align}
Let us define the eigenfunction $G^{\widehat{\Delta}^-_{12}}_{\widehat{\Delta}}(\upsilon)$ through
\begin{align}\label{eq:block exp 2}
\begin{aligned}
        \langle\widehat{\CD}|\,\CR\{ &\widehat{\CO}_{\widehat{\Delta}_1}(Q_1)\,\widehat{\CO}_{\widehat{\Delta}_2}(Q_2)\}\,|\,\widehat{\CO}_{\widehat{\Delta}}\,|\, \CO_{\Delta}(P)\,|\Omega\rangle\\
        &=\frac{b(\CO,\widehat{\CO})\,c(\widehat{\CO},\widehat{\CO}_1,\widehat{\CO}_2)/c(\widehat{\CO},\widehat{\CO})}{(P\circ P)^{\frac{\Delta}{2}} \,(-2P\bullet Q_1)^{\frac{\widehat{\Delta}^-_{12}}{2}}\,(-2P\bullet Q_2)^{\frac{\widehat{\Delta}^-_{21}}{2}}\,(-2Q_1\bullet Q_2)^{\frac{\widehat{\Delta}^+_{12}}{2}} }\,G^{\widehat{\Delta}^-_{12}}_{\widehat{\Delta}}(\upsilon)\ ,
\end{aligned}
\end{align}
where we denoted the bulk-defect two-point coefficient, defect three-point coefficient, and defect two-point coefficient by $b(\CO,\widehat{\CO})$, $c(\widehat{\CO},\widehat{\CO}_1,\widehat{\CO}_2)$ and $c(\widehat{\CO},\widehat{\CO})$ (see \eqref{eq:scalar correlator DCFT 2 and 3}).
Inserting the $\mathrm{SO}(1,p+1)$ Casimir in between $\widehat{\CO}_{\widehat{\Delta}_2}(Q_2)$ and $|\,\widehat{\CO}_{\widehat{\Delta}}\,|$ above and using \eqref{eq:eigenvector of defect conformal group} and \eqref{eq:action of conf gen}, we have
\begin{align}
    \begin{aligned}
 \langle\widehat{\CD}|\,&\CR\{ \widehat{\CO}_{\widehat{\Delta}_1}(Q_1)\,\widehat{\CO}_{\widehat{\Delta}_2}(Q_2)\}\,\left(\frac{1}{2}\,\mathbf{J}^{AB}\mathbf{J}_{AB}\right)\,|\,\widehat{\CO}_{\widehat{\Delta}}\,|\, \CO_{\Delta}(P)\,|\Omega\rangle\\
  &=  -\widehat{\Delta}\,(\widehat{\Delta}-p)\,  \langle\widehat{\CD}|\,\CR\{ \widehat{\CO}_{\widehat{\Delta}_1}(Q_1)\,\widehat{\CO}_{\widehat{\Delta}_2}(Q_2)\}\,|\,\widehat{\CO}_{\widehat{\Delta}}\,|\, \CO_{\Delta}(P)\,|\Omega\rangle\\
  &=    \frac{1}{2}\,\CJ_{AB}(P)\,\CJ^{AB}(P)\,  \langle\widehat{\CD}|\,\CR\{ \widehat{\CO}_{\widehat{\Delta}_1}(Q_1)\,\widehat{\CO}_{\widehat{\Delta}_2}(Q_2)\}\,|\,\widehat{\CO}_{\widehat{\Delta}}\,|\, \CO_{\Delta}(P)\,|\Omega\rangle\ .
    \end{aligned}
\end{align}
Comparing these two expressions, we find
\begin{align}\label{eq:Casimir eq 0}
\begin{aligned}
     & \left[  \frac{1}{2}\,\CJ_{AB}(P)\,\CJ^{AB}(P)+\widehat{\Delta}\,(\widehat{\Delta}-p)\right]\\
  &\qquad\qquad\cdot\frac{G^{\widehat{\Delta}^-_{12}}_{\widehat{\Delta}}(\upsilon)}{(P\circ P)^{\frac{\Delta}{2}} \,(-2P\bullet Q_1)^{\frac{\widehat{\Delta}^-_{12}}{2}}\,(-2P\bullet Q_2)^{\frac{\widehat{\Delta}^-_{21}}{2}}\,(-2Q_1\bullet Q_2)^{\frac{\widehat{\Delta}^+_{12}}{2}} }=0\ .
\end{aligned}
\end{align}
The differential operator commutes with $P\circ P$ and $(-2Q_1\bullet Q_2)$ and \eqref{eq:Casimir eq 0} can be reduced to
\begin{align}\label{eq:Casimir eq 1}
  \left[  \frac{1}{2}\,\CJ_{AB}(P)\,\CJ^{AB}(P)+\widehat{\Delta}\,(\widehat{\Delta}-p)\right]\,\frac{G^{\widehat{\Delta}^-_{12}}_{\widehat{\Delta}}(\upsilon)}{(-2P\bullet Q_1)^{\widehat{\Delta}^-_{12}/2}(-2P\bullet Q_2)^{\widehat{\Delta}^-_{21}/2} }=0\ .
\end{align}
A short calculation shows
\begin{align}\label{eq:Conformal Casimir equation}
\begin{aligned}
    & \left\{4\upsilon^2(1-\upsilon)\partial_\upsilon^2+\left[4(1-\upsilon)-2p\right]\,\upsilon\,\partial_\upsilon+(\widehat{\Delta}^-_{12})^2\,\upsilon-\widehat{\Delta}\,(\widehat{\Delta}-p)\right\}\,G^{\widehat{\Delta}^-_{12}}_{\widehat{\Delta}}(\upsilon)=0\ .
\end{aligned}
\end{align}
After setting $G^{\widehat{\Delta}^-_{12}}_{\widehat{\Delta}}(\upsilon)=\upsilon^{\widehat{\Delta}/2}\,f(\upsilon)$ and some manipulations, it turns out that $f(\upsilon)$ satisfies the following hypergeometric differential equation:
\begin{align}\label{eq:Conformal Casimir equation ast}
    \left\{\upsilon(1-\upsilon)\,\partial_\upsilon^2+\left[\widehat{\Delta}+1-\frac{p}{2}-(\widehat{\Delta}+1)\,\upsilon\,\partial_\upsilon-\frac{\widehat{\Delta}+\widehat{\Delta}^-_{12}}{2}\,\frac{\widehat{\Delta}-\widehat{\Delta}^-_{12}}{2}\right]\right\} \,f(\upsilon)=0\ .
\end{align}
To see the proper boundary condition to this differential equation, consider the DOE of $\CO_{\Delta}$:
\begin{align}
    \CO_{\Delta}(x)\supset\frac{b(\CO,\widehat{\CO})/c(\widehat{\CO},\widehat{\CO})}{|x_\perp|^{\Delta-\widehat{\Delta}}}\,\widehat{\CO}_{\widehat{\Delta}}(\hat{x})\ ,
\end{align}
Taking $\upsilon$ to zero is equivalent to defect OPE limit $|x_\perp|\to0$, where
\begin{align}
\begin{aligned}
    \langle\, \CO_{\Delta}(x)\,\widehat{\CO}_{\widehat{\Delta}_1}&(\hat{y}_1)\,\widehat{\CO}_{\widehat{\Delta}_2}(\hat{y}_2) \,\rangle\xrightarrow[|x_\perp|\to0]{}\sum_{\widehat{\Delta}}\,\frac{b(\CO,\widehat{\CO})/c(\widehat{\CO},\widehat{\CO})}{|x_\perp|^{\Delta-\widehat{\Delta}}}\,\langle\,\widehat{\CO}_{\widehat{\Delta}}(\hat{x})\,\widehat{\CO}_{\widehat{\Delta}_1}(\hat{y}_1)\,\widehat{\CO}_{\widehat{\Delta}_2}(y_2) \,\rangle\\
    &=\sum_{\widehat{\Delta}}\,\frac{b(\CO,\widehat{\CO})\,c(\widehat{\CO},\widehat{\CO}_1,\widehat{\CO}_2)/c(\widehat{\CO},\widehat{\CO})}{|x_\perp|^{\Delta-\widehat{\Delta}}}\,\frac{1}{|\hat{x}-\hat{y}_1|^{\widehat{\Delta}+\widehat{\Delta}^-_{12}}\,|\hat{x}-\hat{y}_2|^{\widehat{\Delta}+\widehat{\Delta}^-_{21}}\,|\hat{y}_{12}|^{\widehat{\Delta}^+_{12}-\widehat{\Delta}}}  \ .
\end{aligned}
\end{align}
Combining this with \eqref{eq:block exp 1} and \eqref{eq:block exp 2} leads
\begin{align}\label{eq:conformal block boundary condition}
    G^{\widehat{\Delta}^-_{12}}_{\widehat{\Delta}}(\upsilon)\xrightarrow[\upsilon\to0]{}\,\upsilon^{\widehat{\Delta}/2}\ .
\end{align}
This boundary condition \eqref{eq:conformal block boundary condition} singles out the proper solution to this differential equation \eqref{eq:Conformal Casimir equation ast} by $f(\upsilon)={}_2F_1\left(\frac{\widehat{\Delta}+\widehat{\Delta}^-_{12}}{2},\frac{\widehat{\Delta}-\widehat{\Delta}^-_{12}}{2};\widehat{\Delta}+1-\frac{p}{2};\upsilon\right)$.

To sum up, the conformal block expansion for the bulk-defect-defect three-point function is
\begin{align}\label{eq:conformal block expansion}
\begin{aligned}
          \langle\, \CO_{\Delta}(x)&\,\widehat{\CO}_{\widehat{\Delta}_1}(\hat{y}_1)\,\widehat{\CO}_{\widehat{\Delta}_2}(y_2) \,\rangle\\
          &=T^{\widehat{\Delta}_1,\widehat{\Delta}_2}_{\Delta}(x,\hat{y}_1,\hat{y_2})\times  \sum_{\widehat{\Delta}}\, \frac{b(\CO,\widehat{\CO})\,c(\widehat{\CO},\widehat{\CO}_1,\widehat{\CO}_2)}{c(\widehat{\CO},\widehat{\CO})}\,G^{\widehat{\Delta}^-_{12}}_{\widehat{\Delta}}(\upsilon)\ ,
\end{aligned}
\end{align}
with $T^{\widehat{\Delta}_1,\widehat{\Delta}_2}_{\Delta}(x,\hat{y}_1,\hat{y_2})$ being some function that transforms covariantly with three-point functions under the residual conformal group $\mathrm{SO}(1,p+1)\times\mathrm{SO}(d-p)$:
\begin{align}
    T^{\widehat{\Delta}_1,\widehat{\Delta}_2}_{\Delta}(x,\hat{y}_1,\hat{y_2})=\frac{1}{|x_\perp|^{\Delta} \,|x-\hat{y}_1|^{\widehat{\Delta}^-_{12}}\,|x-\hat{y}_2|^{\widehat{\Delta}^-_{21}}\,|\hat{y}_{12}|^{\widehat{\Delta}^+_{12}} }\ .
\end{align}
The conformal block $G^{\widehat{\Delta}^-_{12}}_{\widehat{\Delta}}(\upsilon)$ is given by
\begin{align}\label{eq:conformal block}
    G^{\widehat{\Delta}^-_{12}}_{\widehat{\Delta}}(\upsilon)=\upsilon^{\widehat{\Delta}/2}\,{}_2F_1\left(\frac{\widehat{\Delta}+\widehat{\Delta}^-_{12}}{2},\frac{\widehat{\Delta}-\widehat{\Delta}^-_{12}}{2};\widehat{\Delta}+1-\frac{p}{2};\upsilon\right)\ ,
\end{align}
where the cross-ratio $\upsilon$ is
\begin{align}
        \upsilon=\frac{|x_\perp|^2\,|\hat{y}_{12}|^2}{|x-\hat{y}_1|^2\,|x-\hat{y}_2|^2}
    \ .
\end{align}

\subsection{Reconstruction of conformal block from defect operator expansions}\label{app:Reconstruction of conformal block}
We now derive the conformal block expansion \eqref{eq:conformal block expansion} more directly, by summing DOEs for all orders. The calculation is almost parallel to the ones presented in \cite{Ferrara:1973vz} and \cite{Ferrara:1974nf} about conformal four-point functions.

\paragraph{Defect operator expansion.}
The DOEs are completely fixed to reproduce bulk-defect two-point functions with the normalization of the defect two-point functions \eqref{eq:scalar correlator DCFT 2 and 3}. In particular, for the defect scalar channel, the expression reads (see appendix B.1 of \cite{Billo:2016cpy})
\begin{align}\label{eq:scalar btd OPE full}
\begin{aligned}
 \CO_{\Delta}(x)\supset\sum_{\widehat{\CO}} \,\frac{b(\CO,\widehat{\CO})/c(\widehat{\CO},\widehat{\CO})}{|x_\perp|^{\Delta-\widehat{\Delta}}}\,\sum_{n=0}^\infty\,\frac{(-1)^n\,|x_\perp|^{2n}}{2^{2n}\,(\widehat{\Delta}+1-p/2)_n\,n!}\, (\widehat{\partial}^{\,2}_x)^n\,\widehat{\CO}_{\widehat{\Delta}}(\hat{x})\ ,
\end{aligned}
\end{align}
with $\widehat{\partial}^{\,2}_x=\partial^2/\partial \hat{x}^a\partial \hat{x}_a$.
One can check the validity of this expansion by applying it inside the bulk-to-defect two-point functions:\footnote{The following two formulas are used;
\begin{align*}
    (\widehat{\partial}^{\,2}_x)^n\,\frac{1}{|\hat{x}-\hat{y}|^{2\widehat{\Delta}}}=\frac{2^{2n}\,(\widehat{\Delta})_n\,(\widehat{\Delta}+1-p/2)_n}{|\hat{x}-\hat{y}|^{2\widehat{\Delta}+2n}}\ ,\qquad   \sum_{n=0}^\infty\,\frac{(\widehat{\Delta})_n}{n!}\, \left(-\frac{|x_\perp|^{2}}{|\hat{x}-\hat{y}|^{2}}\right)^n=\frac{1}{(1+|x_\perp|^{2}/|\hat{x}-\hat{y}|^{2})^{\widehat{\Delta}}}\ .
\end{align*}
}
\begin{align}
\begin{aligned}
  \langle\,\CO_{\Delta}(x)\,&\widehat{\CO}_{\widehat{\Delta}}(\hat{y})\,\rangle\\
  &=\frac{b(\CO,\widehat{\CO})/c(\widehat{\CO},\widehat{\CO})}{|x_\perp|^{\Delta-\widehat{\Delta}}}\,\sum_{n=0}^\infty\,\frac{(-1)^n\,|x_\perp|^{2n}}{2^{2n}\,(\widehat{\Delta}+1-p/2)_n\,n!}\, (\widehat{\partial}^{\,2}_x)^n\,\langle\,\widehat{\CO}_{\widehat{\Delta}}(\hat{x})\,\widehat{\CO}_{\widehat{\Delta}}(\hat{y})\,\rangle\\
  &=\frac{b(\CO,\widehat{\CO})}{|x_\perp|^{\Delta-\widehat{\Delta}}}\,\sum_{n=0}^\infty\,\frac{(-1)^n\,|x_\perp|^{2n}}{2^{2n}\,(\widehat{\Delta}+1-p/2)_n\,n!}\, (\widehat{\partial}^{\,2}_x)^n\,\frac{1}{|\hat{x}-\hat{y}|^{2\widehat{\Delta}}}\\
    &=\frac{b(\CO,\widehat{\CO})}{|x_\perp|^{\Delta-\widehat{\Delta}}\,|\hat{x}-\hat{y}|^{2\widehat{\Delta}}}\,\sum_{n=0}^\infty\,\frac{(\widehat{\Delta})_n}{n!}\, \left(-\frac{|x_\perp|^{2}}{|\hat{x}-\hat{y}|^{2}}\right)^n\\
    &=\frac{b(\CO,\widehat{\CO})}{(|\hat{x}-\hat{y}|^2+|x_\perp|^2)^{\widehat{\Delta}}\,|x_\perp|^{\Delta-\widehat{\Delta}}}\ .
\end{aligned}
\end{align}
By considering the Fourier transformation of $\widehat{\CO}_{\widehat{\Delta}}(\hat{x})$ defined through the following relation:
\begin{align}
\widehat{\CO}_{\widehat{\Delta}}(\hat{x})=\int\frac{\d^p \hat{q}}{(2\pi)^p}\, e^{\i\,\hat{q}\cdot \hat{x}}\,\widehat{\CO}_{\widehat{\Delta}}(\hat{q}) \ ,
\end{align}
we obtain an alternative expression of the DOE:
\begin{align}\label{eq:momentum space expression btd}
\begin{aligned}
 \CO_{\Delta}(x)&\supset\sum_{\widehat{\CO}} \,\frac{b(\CO,\widehat{\CO})/c(\widehat{\CO},\widehat{\CO})}{|x_\perp|^{\Delta-\widehat{\Delta}}}\,\Gamma(\widehat{\Delta}+1-p/2)\,\left(\frac{|x_\perp|}{2}\right)^{p/2-\widehat{\Delta}} \\
 &\qquad\qquad\times \int\,\frac{\d^p \hat{q}}{(2\pi)^p}\, e^{\i\,\hat{q}\cdot \hat{x}}\,|\hat{q}|^{p/2-\widehat{\Delta}}\, I_{\widehat{\Delta}-p/2}(|x_\perp|\,|\hat{q}|) \,\widehat{\CO}_{\widehat{\Delta}}(\hat{q})\ ,
\end{aligned}
\end{align}
Here, $I_\nu(z)$ is the modified Bessel function $I_\nu(z)=\sum_{n=0}^{\infty}\,\frac{1}{\Gamma(\nu+n+1)\,n!}\,\left(\frac{z}{2}\right)^{\nu+2n}$.

\paragraph{Summing over defect operator expansions for all orders.}
Let us apply the DOE inside bulk-defect-defect three-point function using \eqref{eq:momentum space expression btd}:
\begin{align}
\begin{aligned}
     & \langle\, \CO_{\Delta}(x)\,\widehat{\CO}_{\widehat{\Delta}_1}(\hat{y}_1)\,\widehat{\CO}_{\widehat{\Delta}_2}(\hat{y}_2) \,\rangle
 = \sum_{\widehat{\CO}} \,\frac{b(\CO,\widehat{\CO})/c(\widehat{\CO},\widehat{\CO})}{2^{p/2-\widehat{\Delta}}\,|x_\perp|^{\Delta-p/2}}\cdot \Gamma(\widehat{\Delta}+1-p/2)\\
      &\qquad\cdot\int\frac{\d^p \hat{q}}{(2\pi)^p}\, e^{\i\,\hat{q}\cdot \hat{x}}\,|\hat{q}|^{p/2-\widehat{\Delta}}\, I_{\widehat{\Delta}-p/2}(|x_\perp|\,|\hat{q}|) \,\langle\, \widehat{\CO}_{\widehat{\Delta}}(\hat{q})\,\widehat{\CO}_{\widehat{\Delta}_1}(\hat{y}_1)\,\widehat{\CO}_{\widehat{\Delta}_2}(\hat{y}_2) \,\rangle\ .
\end{aligned}
\end{align}
The partial Fourier transformation of the three-point function is given by (see e.g., \cite[equation (3.41)]{Fradkin:1996is})
{\small\begin{align}
\begin{aligned}
 \langle\, &\widehat{\CO}_{\widehat{\Delta}}(\hat{q})\,\widehat{\CO}_{\widehat{\Delta}_1}(\hat{y}_1)\,\widehat{\CO}_{\widehat{\Delta}_2}(\hat{y}_2) \,\rangle=\int\d^p \hat{x}\,e^{-\i\,\hat{q}\cdot\hat{x}}\,\langle\, \widehat{\CO}_{\widehat{\Delta}}(\hat{x})\,\widehat{\CO}_{\widehat{\Delta}_1}(\hat{y}_1)\,\widehat{\CO}_{\widehat{\Delta}_2}(\hat{y}_2) \,\rangle\\
&=\frac{c(\widehat{\CO},\widehat{\CO}_1,\widehat{\CO}_2)}{|\hat{y}_{12}|^{\Delta^+_{12}-\widehat{\Delta}}}\,\frac{2\,\pi^{p/2}}{\Gamma\left(\frac{\widehat{\Delta}+\widehat{\Delta}^-_{12}}{2}\right)\Gamma\left(\frac{\widehat{\Delta}+\widehat{\Delta}^-_{21}}{2}\right)}\,\left(\frac{|\hat{q}|}{2\,|\hat{y}_{12}|}\right)^{\widehat{\Delta}-p/2}\\
&\qquad\cdot \int_0^1\d\xi\,\xi^{\frac{\widehat{\Delta}^-_{12}+p/2}{2}-1}\,(1-\xi)^{\frac{\widehat{\Delta}^-_{21}+p/2}{2}-1}\, e^{-\i\,\hat{q}\cdot [\xi\,\hat{y}_1+(1-\xi)\,\hat{y}_2]}\,K_{\widehat{\Delta}-p/2}\left(\sqrt{\xi(1-\xi)}\,|\hat{q}|\,|\hat{y}_{12}|\right)\ .
\end{aligned}
\end{align}}
Hence,
{\small\begin{align}\label{eq:three point explicit cal 1}
\begin{aligned}
     & \langle\, \CO_{\Delta}(x)\,\widehat{\CO}_{\widehat{\Delta}_1}(\hat{y}_1)\,\widehat{\CO}_{\widehat{\Delta}_2}(\hat{y}_2) \,\rangle \\
     &= \sum_{\widehat{\CO}} \,\frac{b(\CO,\widehat{\CO})\,c(\widehat{\CO},\widehat{\CO}_1,\widehat{\CO}_2)}{c(\widehat{\CO},\widehat{\CO})}\\
      &\qquad\cdot \frac{2\,\pi^{p/2}\,\Gamma(\widehat{\Delta}+1-p/2)}{|\hat{y}_{12}|^{\widehat{\Delta}^+_{12}-p/2}\,|x_\perp|^{\Delta-p/2}\,\Gamma\left(\frac{\widehat{\Delta}+\widehat{\Delta}^-_{12}}{2}\right)\Gamma\left(\frac{\widehat{\Delta}+\widehat{\Delta}^-_{21}}{2}\right)}\cdot \int_0^1\d\xi\,\xi^{\frac{\widehat{\Delta}^-_{12}+p/2}{2}-1}\,(1-\xi)^{\frac{\widehat{\Delta}^-_{21}+p/2}{2}-1}\\
      &\qquad\cdot\int\frac{\d^p \hat{q}}{(2\pi)^p}\,e^{\i\,\hat{q}\cdot [\xi(\hat{x}-\hat{y}_1)+(1-\xi)(\hat{x}-\hat{y}_2)]}\, I_{\widehat{\Delta}-p/2}(|x_\perp|\,|\hat{q}|) \,K_{\widehat{\Delta}-p/2}\left(\sqrt{\xi(1-\xi)}\,|\hat{q}|\,|\hat{y}_{12}|\right)\ .
\end{aligned}
\end{align}}
We first perform the angular part of the $\hat{q}$-integral as follows\footnote{The relevant formula is $\int_0^\pi\,\d\theta\,(\sin\theta)^a\,e^{-\i\,b\,\cos\theta}=\int_0^\pi\,\d\theta\,(\sin\theta)^a\,\cos(b\cos\theta)=\left(\frac{b}{2}\right)^{-a/2}\sqrt{\pi}\,\Gamma(a/2+1/2)\,J_{a/2}(b)$. Also, $\mathrm{Vol}\,(\mathbb{S}^n)=\frac{2\,\pi^{\frac{n+1}{2}}}{\Gamma\left(\frac{n+1}{2}\right)}$ is the volume of unit $n$-sphere.}
\begin{align}
\begin{aligned}
     \int\frac{\d^p \hat{q}}{(2\pi)^p}\, e^{-\i\,\hat{q}\cdot\hat{x}}\cdots&=\mathrm{Vol}\,(\mathbb{S}^{p-2})\,\int_0^\infty \d |\hat{q}|\, |\hat{q}|^{p-1}\, \int_0^\pi \d\theta \,(\sin\theta)^{p-2}\,e^{-\i\,|\hat{q}| \,|\hat{x}|\,\cos\theta}\cdots\\
     &=\frac{1}{|\hat{x}|^{p/2-1}}\cdot \int_0^\infty\frac{\d |\hat{q}|}{(2\pi)^{p/2}}\, |\hat{q}|^{p/2}\, J_{p/2-1}(|\hat{x}|\,|\hat{q}|)\cdots\ .
\end{aligned}
\end{align}
Then, the remaining $\hat{q}$-integral in the last line of \eqref{eq:three point explicit cal 1} can be integrated to give a hypergeometric function:\footnote{The following integration formula is used (see e.g., \cite[equation (6.578.11)]{zwillinger2014table})
\begin{align*}
    \int_0^\infty\d x\, x^{\nu+1}\,K_{\mu}(ax)\,I_{\mu}(bx)\,J_{\nu}(cx)=\frac{\Gamma(\mu+\nu)}{2^{\mu+1}\,\Gamma(\mu+1)}\cdot\frac{c^{\nu}}{(ab)^{\nu+1}\,u^{\mu+\nu+1}}\,{}_2F_1\left(\frac{\mu+\nu+1}{2},\frac{\mu+\nu+2}{2};\mu+1;\frac{1}{u^2}\right)\ ,
\end{align*}
with $2ab u=a^2+b^2+c^2$.
}
\begin{align}
\begin{aligned}
&(\text{The last line of \eqref{eq:three point explicit cal 1}})\\
  &= \frac{\Gamma(\widehat{\Delta})\,[\xi(1-\xi)]^{-\frac{p/2}{2}}}{2^{\widehat{\Delta}+1}\,\pi^{p/2}\,\Gamma(\widehat{\Delta}-p/2+1)\,|x_\perp|^{p/2}\,|\hat{y}_{12}|^{p/2}}\, \frac{1}{\tilde{u}^{\widehat{\Delta}}}\,{}_2F_1\left(\frac{\widehat{\Delta}}{2},\frac{\widehat{\Delta}+1}{2};\widehat{\Delta}-p/2+1;\frac{1}{\tilde{u}^2}\right) \ .
\end{aligned}
\end{align}
Here we defined $\tilde{u}$ by
\begin{align}
    \tilde{u}=\frac{|x_\perp|^2+|\hat{x}-\hat{y}_1|^2\,\xi+|\hat{x}-\hat{y}_2|^2\,(1-\xi)}{2\,|x_\perp|\,|\hat{y}_{12}|\,\sqrt{\xi(1-\xi)}} \ .
\end{align}
Plugging this expression into the last two lines of \eqref{eq:three point explicit cal 1} and using a Mellin–Barnes integral representation of the hypergeometric function
\begin{align}\label{eq:Mellin 2F1}
{}_2F_1(a,b;c;z)=\frac{\Gamma(c)}{\Gamma(a)\Gamma(b)}\,\int_{-\i\,\infty}^{\i\,\infty}\frac{\d s}{2\pi\i}\,\frac{\Gamma(a+s)\Gamma(b+s)\Gamma(-s)}{\Gamma(c+s)}\,(-z)^s\ ,
\end{align}
we have
{\small\begin{align}\label{eq:geodesic integral}
\begin{aligned}
&(\text{The last two lines of \eqref{eq:three point explicit cal 1}})\\
          &=  T^{\widehat{\Delta}_1,\widehat{\Delta}_2}_{\Delta}(x,\hat{y}_1,\hat{y_2})\,\upsilon^{\widehat{\Delta}}\\
  &\qquad\cdot \frac{\Gamma(\widehat{\Delta}-p/2+1)}{\Gamma\left(\frac{\widehat{\Delta}+\widehat{\Delta}^-_{12}}{2}\right)\Gamma\left(\frac{\widehat{\Delta}+\widehat{\Delta}^-_{21}}{2}\right)}\, \int_{-\i\,\infty}^{\i\,\infty}\frac{\d s}{2\pi\i}\,\frac{\Gamma\left(\frac{\widehat{\Delta}+\widehat{\Delta}^-_{12}}{2}+s\right)\Gamma\left(\frac{\widehat{\Delta}+\widehat{\Delta}^-_{21}}{2}+s\right)\Gamma(-s)}{\Gamma(\widehat{\Delta}-p/2+s+1)}\,(-\upsilon)^s\ ,
\end{aligned}
\end{align}}
where we have implemented the following relation to perform the $\xi$-integral
\begin{align}
    \int_0^1\d x\, \frac{x^{\mu-1}(1-x)^{\nu-1}}{[ax+b\,(1-x)+c]^{\mu+\nu}}&=\frac{\Gamma(\mu)\,\Gamma(\nu)}{(a+c)^\mu (b+c)^\nu\,\Gamma(\mu+\nu)}\label{eq:2F1 application}\ .
\end{align}
The last line of \eqref{eq:geodesic integral} can be identified as the Mellin–Barnes integral representation of hypergeometric function \eqref{eq:Mellin 2F1} and we end up with the conformal block expansion \eqref{eq:conformal block expansion}.

 \bibliographystyle{JHEP}
 \bibliography{DCFT}

\providecommand{\href}[2]{#2}\begingroup\raggedright\begin{thebibliography}{10}

\bibitem{Wilson:1974sk}
K.~G. Wilson, \emph{{Confinement of Quarks}},
  \href{https://doi.org/10.1103/PhysRevD.10.2445}{\emph{Phys. Rev. D}
  {\bfseries 10} (1974) 2445}.

\bibitem{tHooft:1977nqb}
G.~'t~Hooft, \emph{{On the Phase Transition Towards Permanent Quark
  Confinement}},
  \href{https://doi.org/10.1016/0550-3213(78)90153-0}{\emph{Nucl. Phys. B}
  {\bfseries 138} (1978) 1}.

\bibitem{Kondo:1964}
J.~Kondo, \emph{{Resistance Minimum in Dilute Magnetic Alloys}},
  \href{https://doi.org/10.1143/PTP.32.37}{\emph{Progress of Theoretical
  Physics} {\bfseries 32} (1964) 37}
  [\href{https://arxiv.org/abs/https://doi.org/10.1143/PTP.32.37}{{\ttfamily
  https://doi.org/10.1143/PTP.32.37}}].

\bibitem{LudwigAffleck:1991}
A.~W.~W. Ludwig and I.~Affleck, \emph{Exact, asymptotic, three-dimensional,
  space- and time-dependent, green's functions in the multichannel kondo
  effect}, \href{https://doi.org/10.1103/PhysRevLett.67.3160}{\emph{Phys. Rev.
  Lett.} {\bfseries 67} (1991) 3160}.

\bibitem{Affleck:1995ge}
I.~Affleck, \emph{{Conformal field theory approach to the Kondo effect}},
  {\emph{Acta Phys. Polon. B} {\bfseries 26} (1995) 1869}
  [\href{https://arxiv.org/abs/cond-mat/9512099}{{\ttfamily
  cond-mat/9512099}}].

\bibitem{Allais:2014fqa}
A.~Allais and S.~Sachdev, \emph{{Spectral function of a localized fermion
  coupled to the Wilson-Fisher conformal field theory}},
  \href{https://doi.org/10.1103/PhysRevB.90.035131}{\emph{Phys. Rev. B}
  {\bfseries 90} (2014) 035131}
  [\href{https://arxiv.org/abs/1406.3022}{{\ttfamily 1406.3022}}].

\bibitem{Cuomo:2021kfm}
G.~Cuomo, Z.~Komargodski and M.~Mezei, \emph{{Localized magnetic field in the
  O(N) model}}, \href{https://doi.org/10.1007/JHEP02(2022)134}{\emph{JHEP}
  {\bfseries 02} (2022) 134}
  [\href{https://arxiv.org/abs/2112.10634}{{\ttfamily 2112.10634}}].

\bibitem{Rattazzi:2008pe}
R.~Rattazzi, V.~S. Rychkov, E.~Tonni and A.~Vichi, \emph{{Bounding scalar
  operator dimensions in 4D CFT}},
  \href{https://doi.org/10.1088/1126-6708/2008/12/031}{\emph{JHEP} {\bfseries
  12} (2008) 031} [\href{https://arxiv.org/abs/0807.0004}{{\ttfamily
  0807.0004}}].

\bibitem{Poland:2018epd}
D.~Poland, S.~Rychkov and A.~Vichi, \emph{{The Conformal Bootstrap: Theory,
  Numerical Techniques, and Applications}},
  \href{https://doi.org/10.1103/RevModPhys.91.015002}{\emph{Rev. Mod. Phys.}
  {\bfseries 91} (2019) 015002}
  [\href{https://arxiv.org/abs/1805.04405}{{\ttfamily 1805.04405}}].

\bibitem{Liendo:2012hy}
P.~Liendo, L.~Rastelli and B.~C. van Rees, \emph{{The Bootstrap Program for
  Boundary CFT$_d$}},
  \href{https://doi.org/10.1007/JHEP07(2013)113}{\emph{JHEP} {\bfseries 07}
  (2013) 113} [\href{https://arxiv.org/abs/1210.4258}{{\ttfamily 1210.4258}}].

\bibitem{Bissi:2018mcq}
A.~Bissi, T.~Hansen and A.~S\"oderberg, \emph{{Analytic Bootstrap for Boundary
  CFT}}, \href{https://doi.org/10.1007/JHEP01(2019)010}{\emph{JHEP} {\bfseries
  01} (2019) 010} [\href{https://arxiv.org/abs/1808.08155}{{\ttfamily
  1808.08155}}].

\bibitem{Dey:2020jlc}
P.~Dey and A.~S\"oderberg, \emph{{On analytic bootstrap for interface and
  boundary CFT}}, \href{https://doi.org/10.1007/JHEP07(2021)013}{\emph{JHEP}
  {\bfseries 07} (2021) 013}
  [\href{https://arxiv.org/abs/2012.11344}{{\ttfamily 2012.11344}}].

\bibitem{Padayasi:2021sik}
J.~Padayasi, A.~Krishnan, M.~A. Metlitski, I.~A. Gruzberg and M.~Meineri,
  \emph{{The extraordinary boundary transition in the 3d O(N) model via
  conformal bootstrap}},
  \href{https://doi.org/10.21468/SciPostPhys.12.6.190}{\emph{SciPost Phys.}
  {\bfseries 12} (2022) 190}
  [\href{https://arxiv.org/abs/2111.03071}{{\ttfamily 2111.03071}}].

\bibitem{Gliozzi:2015qsa}
F.~Gliozzi, P.~Liendo, M.~Meineri and A.~Rago, \emph{{Boundary and Interface
  CFTs from the Conformal Bootstrap}},
  \href{https://doi.org/10.1007/JHEP05(2015)036}{\emph{JHEP} {\bfseries 05}
  (2015) 036} [\href{https://arxiv.org/abs/1502.07217}{{\ttfamily
  1502.07217}}].

\bibitem{Gimenez-Grau:2022ebb}
A.~Gimenez-Grau, \emph{{Probing magnetic line defects with two-point
  functions}},  \href{https://arxiv.org/abs/2212.02520}{{\ttfamily
  2212.02520}}.

\bibitem{Gimenez-Grau:2022czc}
A.~Gimenez-Grau, E.~Lauria, P.~Liendo and P.~van Vliet, \emph{{Bootstrapping
  line defects with $O(2)$ global symmetry}},
  \href{https://arxiv.org/abs/2208.11715}{{\ttfamily 2208.11715}}.

\bibitem{Bianchi:2022sbz}
L.~Bianchi, D.~Bonomi and E.~de~Sabbata, \emph{{Analytic bootstrap for the
  localized magnetic field}},
  \href{https://arxiv.org/abs/2212.02524}{{\ttfamily 2212.02524}}.

\bibitem{Rychkov:2015naa}
S.~Rychkov and Z.~M. Tan, \emph{{The $\epsilon$-expansion from conformal field
  theory}}, \href{https://doi.org/10.1088/1751-8113/48/29/29FT01}{\emph{J.
  Phys. A} {\bfseries 48} (2015) 29FT01}
  [\href{https://arxiv.org/abs/1505.00963}{{\ttfamily 1505.00963}}].

\bibitem{Basu:2015gpa}
P.~Basu and C.~Krishnan, \emph{{$\epsilon$-expansions near three dimensions
  from conformal field theory}},
  \href{https://doi.org/10.1007/JHEP11(2015)040}{\emph{JHEP} {\bfseries 11}
  (2015) 040} [\href{https://arxiv.org/abs/1506.06616}{{\ttfamily
  1506.06616}}].

\bibitem{Ghosh:2015opa}
S.~Ghosh, R.~K. Gupta, K.~Jaswin and A.~A. Nizami, \emph{{$\epsilon$-Expansion
  in the Gross-Neveu model from conformal field theory}},
  \href{https://doi.org/10.1007/JHEP03(2016)174}{\emph{JHEP} {\bfseries 03}
  (2016) 174} [\href{https://arxiv.org/abs/1510.04887}{{\ttfamily
  1510.04887}}].

\bibitem{Raju:2015fza}
A.~Raju, \emph{{$\epsilon$-Expansion in the Gross-Neveu CFT}},
  \href{https://doi.org/10.1007/JHEP10(2016)097}{\emph{JHEP} {\bfseries 10}
  (2016) 097} [\href{https://arxiv.org/abs/1510.05287}{{\ttfamily
  1510.05287}}].

\bibitem{Nii:2016lpa}
K.~Nii, \emph{{Classical equation of motion and Anomalous dimensions at leading
  order}}, \href{https://doi.org/10.1007/JHEP07(2016)107}{\emph{JHEP}
  {\bfseries 07} (2016) 107}
  [\href{https://arxiv.org/abs/1605.08868}{{\ttfamily 1605.08868}}].

\bibitem{Giombi:2017rhm}
S.~Giombi, V.~Kirilin and E.~Skvortsov, \emph{{Notes on Spinning Operators in
  Fermionic CFT}}, \href{https://doi.org/10.1007/JHEP05(2017)041}{\emph{JHEP}
  {\bfseries 05} (2017) 041}
  [\href{https://arxiv.org/abs/1701.06997}{{\ttfamily 1701.06997}}].

\bibitem{Yamaguchi:2016pbj}
S.~Yamaguchi, \emph{{The \ensuremath{\epsilon}-expansion of the codimension two
  twist defect from conformal field theory}},
  \href{https://doi.org/10.1093/ptep/ptw115}{\emph{PTEP} {\bfseries 2016}
  (2016) 091B01} [\href{https://arxiv.org/abs/1607.05551}{{\ttfamily
  1607.05551}}].

\bibitem{Giombi:2020rmc}
S.~Giombi and H.~Khanchandani, \emph{{CFT in AdS and boundary RG flows}},
  \href{https://doi.org/10.1007/JHEP11(2020)118}{\emph{JHEP} {\bfseries 11}
  (2020) 118} [\href{https://arxiv.org/abs/2007.04955}{{\ttfamily
  2007.04955}}].

\bibitem{Soderberg:2017oaa}
A.~S\"oderberg, \emph{{Anomalous Dimensions in the WF O($N$) Model with a
  Monodromy Line Defect}},
  \href{https://doi.org/10.1007/JHEP03(2018)058}{\emph{JHEP} {\bfseries 03}
  (2018) 058} [\href{https://arxiv.org/abs/1706.02414}{{\ttfamily
  1706.02414}}].

\bibitem{Herzog:2022jlx}
C.~P. Herzog and V.~Schaub, \emph{{Fermions in Boundary Conformal Field Theory
  : Crossing Symmetry and $\epsilon$-Expansion}},
  \href{https://arxiv.org/abs/2209.05511}{{\ttfamily 2209.05511}}.

\bibitem{Giombi:2022vnz}
S.~Giombi, E.~Helfenberger and H.~Khanchandani, \emph{{Line Defects in
  Fermionic CFTs}},  \href{https://arxiv.org/abs/2211.11073}{{\ttfamily
  2211.11073}}.

\bibitem{Billo:2016cpy}
M.~Bill\`o, V.~Gon\c{c}alves, E.~Lauria and M.~Meineri, \emph{{Defects in
  conformal field theory}},
  \href{https://doi.org/10.1007/JHEP04(2016)091}{\emph{JHEP} {\bfseries 04}
  (2016) 091} [\href{https://arxiv.org/abs/1601.02883}{{\ttfamily
  1601.02883}}].

\bibitem{Gadde:2016fbj}
A.~Gadde, \emph{{Conformal constraints on defects}},
  \href{https://doi.org/10.1007/JHEP01(2020)038}{\emph{JHEP} {\bfseries 01}
  (2020) 038} [\href{https://arxiv.org/abs/1602.06354}{{\ttfamily
  1602.06354}}].

\bibitem{Kapustin:2005py}
A.~Kapustin, \emph{{Wilson-'t Hooft operators in four-dimensional gauge
  theories and S-duality}},
  \href{https://doi.org/10.1103/PhysRevD.74.025005}{\emph{Phys. Rev. D}
  {\bfseries 74} (2006) 025005}
  [\href{https://arxiv.org/abs/hep-th/0501015}{{\ttfamily hep-th/0501015}}].

\bibitem{luke1969special}
Y.~Luke, \emph{The Special Functions and Their Approximations}, no.~volume 2 in
  Mathematics in Science and Engineering. Academic Press, 1969.

\bibitem{Lemos:2017vnx}
M.~Lemos, P.~Liendo, M.~Meineri and S.~Sarkar, \emph{{Universality at large
  transverse spin in defect CFT}},
  \href{https://doi.org/10.1007/JHEP09(2018)091}{\emph{JHEP} {\bfseries 09}
  (2018) 091} [\href{https://arxiv.org/abs/1712.08185}{{\ttfamily
  1712.08185}}].

\bibitem{BCFTpaper}
T.~Nishioka, Y.~Okuyama and S.~Shimamori. {\it Comments on epsilon expansion of
  the O$(N)$ model with boundary}, in preparation.

\bibitem{Lauria:2020emq}
E.~Lauria, P.~Liendo, B.~C. Van~Rees and X.~Zhao, \emph{{Line and surface
  defects for the free scalar field}},
  \href{https://doi.org/10.1007/JHEP01(2021)060}{\emph{JHEP} {\bfseries 01}
  (2021) 060} [\href{https://arxiv.org/abs/2005.02413}{{\ttfamily
  2005.02413}}].

\bibitem{Buric:2020zea}
I.~Buri\'c and V.~Schomerus, \emph{{Defect Conformal Blocks from Appell
  Functions}}, \href{https://doi.org/10.1007/JHEP05(2021)007}{\emph{JHEP}
  {\bfseries 05} (2021) 007}
  [\href{https://arxiv.org/abs/2012.12489}{{\ttfamily 2012.12489}}].

\bibitem{Karch:2018uft}
A.~Karch and Y.~Sato, \emph{{Conformal Manifolds with Boundaries or Defects}},
  \href{https://doi.org/10.1007/JHEP07(2018)156}{\emph{JHEP} {\bfseries 07}
  (2018) 156} [\href{https://arxiv.org/abs/1805.10427}{{\ttfamily
  1805.10427}}].

\bibitem{Costa:2011mg}
M.~S. Costa, J.~Penedones, D.~Poland and S.~Rychkov, \emph{{Spinning Conformal
  Correlators}}, \href{https://doi.org/10.1007/JHEP11(2011)071}{\emph{JHEP}
  {\bfseries 11} (2011) 071} [\href{https://arxiv.org/abs/1107.3554}{{\ttfamily
  1107.3554}}].

\bibitem{Ferrara:1973vz}
S.~Ferrara, A.~F. Grillo, G.~Parisi and R.~Gatto, \emph{{Covariant expansion of
  the conformal four-point function}},
  \href{https://doi.org/10.1016/0550-3213(72)90587-1,
  10.1016/0550-3213(73)90467-7}{\emph{Nucl. Phys.} {\bfseries B49} (1972) 77}.

\bibitem{Ferrara:1974nf}
S.~Ferrara, A.~F. Grillo, R.~Gatto and G.~Parisi, \emph{{Analyticity properties
  and asymptotic expansions of conformal covariant green's functions}},
  \href{https://doi.org/10.1007/BF02813413}{\emph{Nuovo Cim. A} {\bfseries 19}
  (1974) 667}.

\bibitem{Fradkin:1996is}
E.~Fradkin and M.~Palchik, \emph{{Conformal quantum field theory in
  D-dimensions}}. 1996.

\bibitem{zwillinger2014table}
D.~Zwillinger, \emph{Table of Integrals, Series, and Products}. Elsevier
  Science, 2014.

\end{thebibliography}\endgroup

\end{document}